\documentclass[12pt,letterpaper]{JHEP3}
\usepackage{amsmath,amssymb,scalefnt}
\def\al{\alpha} 
\def\ga{\gamma}

\def\ep{\epsilon}

\def\et{\eta}
\def\th{\theta}

\def\ka{\kappa}
\def\la{\lambda}

\def\si{\sigma}

\def\De{\Delta}

\def\La{\Lambda}

\def\Om{\Omega}

\newcommand{\ben}{\begin{equation}}
\newcommand{\een}{\end{equation}}
\newcommand{\bea}{\begin{eqnarray}}
\newcommand{\eea}{\end{eqnarray}}
\newcommand{\ba}{\begin{array}}
\newcommand{\ea}{\end{array}}
\newcommand{\bit}{\begin{itemize}}
\newcommand{\eit}{\end{itemize}}

\def\math{\mathsurround 0pt}
\def\oversim#1#2{\lower.5pt\vbox{\baselineskip0pt \lineskip-.5pt
 \ialign{$\math#1\hfil##\hfil$\crcr#2\crcr{\scriptstyle\sim}\crcr}}}
\def\lap{\mathrel{\mathpalette\oversim {\scriptstyle <}}}

\def\pa{\partial}

\def\half{\frac{1}{2}}

\newcommand{\vev}[1]{\left\langle#1\right\rangle}

\newcommand{\tenfun}{B}
\newcommand{\powspec}{\mathcal{P}}

\newcommand{\mpl}{m_\mathrm{P}}

\newcommand{\eqn}[1]{Eq.~(\ref{#1})}

\newcommand{\reference}[1]{Ref.~\cite{#1}}

\newcommand{\abbrev}{\scalefont{.9}}

\newcommand{\sm}{{\abbrev SM}}

\newcommand{\ckm}{{\abbrev CKM}}
\newcommand{\cp}{{\abbrev CP}}
\newcommand{\lsp}{{\abbrev LSP}}
\newcommand{\nlsp}{{\abbrev NLSP}}
\newcommand{\mssm}{{\abbrev MSSM}}
\newcommand{\nmssm}{{\abbrev NMSSM}}

\newcommand{\cmssm}{{\abbrev CMSSM}}
\newcommand{\amsb}{{\abbrev AMSB}}

\def\ee{\end{equation}}
\def\Tr{{\rm Tr }}

\def\frak#1#2{{\textstyle{\frac{#1}{#2}}}}

\def\mtilde{\tilde m{}}

\def\GeV{\hbox{GeV}}
\def\TeV{\hbox{TeV}}
\def\half{\frac{1}{2}}

\def\vev#1{\mathopen\langle #1\mathclose\rangle }
\def\nn{\nonumber\\}
\def\DRED{\ifmmode{{\rm DRED}} \else{{DRED}} \fi}
\def\DREDD{\ifmmode{{\rm DRED}'} \else{${\rm DRED}'$} \fi}
\def\NSVZ{\ifmmode{{\rm NSVZ}} \else{{NSVZ}} \fi}

\def\pa{\partial}
\def\ga{\gamma} 
 
\def\ep{\epsilon}
\def \la{\lambda}
\def \La{\Lambda}
\def \th{\theta}
\def\sic{supersymmetric}

\def\sy{supersymmetry}
\def\sic{supersymmetric}

\def\phib{\overline{\phi}}
\def\Phib{\overline\Phi}

\def\mtilde{{\tilde m}}
\author{A.~Basb\o ll$^{1}$, M.~Hindmarsh$^{1}$, 
D.R.T.~Jones$^{2}$\\
$^{1}$Department of Physics and Astronomy, University of Sussex, 
Brighton, East Sussex BN1 9QH,\\
$^{2}$Dept. of Mathematical Sciences,
University of Liverpool, Liverpool L69 3BX, UK} 
\title{Anomaly Mediation and Cosmology}

\abstract{
We consider an extension of the \mssm\ wherein anomaly mediation is the
source of \sy-breaking, and the tachyonic slepton problem is solved by a
gauged $U(1)$ symmetry, which is broken at high energies in a manner 
preserving supersymmetry, thereby also facilitating the see-saw
mechanism for neutrino masses and a natural source for the Higgs
$\mu$-term.  We show that these favourable outcomes can occur both  in the
presence and the absence of a large Fayet-Iliopoulos (FI) $D$-term
associated with the  new $U(1)$. We explore the cosmological
consequences of the model, showing that it  naturally produces a period
of hybrid inflation, terminating in the production of cosmic strings. In
spite of the presence of a $U(1)$ (even with an FI term), inflation is
effected by the $F$-term,  with a $D$-flat tree potential (the FI term,
if present, being cancelled by non-zero  squark and slepton fields).
Calculating the 1-loop corrections to the inflaton potential, we
estimate  the constraints on the parameters of the model from Cosmic
Microwave Background data.  We will see that a consequence of these
constraints is that the Higgs $\mu$-term necessarily small. 
We briefly discuss the mechanisms for
baryogenesis via conventional leptogenesis, the out-of-equilibrium
production of neutrinos from the cosmic strings, or the Affleck-Dine
mechanism. Cosmic string decays also boost the relic density of dark
matter above the low value normally obtained in \amsb{} scenarios.
}

\preprint{
LTH 903\\}
\begin{document}

\section{Introduction}

Low energy \sy\ remains a popular possibility for physics Beyond the Standard 
Model awaiting discovery at the LHC. Much of the associated 
work has concentrated on the \cmssm{} scenario, where it is
assumed that the unification of gauge couplings at high energies is
accompanied by a corresponding unification in both the soft
\sy-breaking scalar masses and the gaugino masses; and also that the
cubic scalar interactions are of the same form as the Yukawa couplings
and related to them by a common constant of proportionality, the
$A$-parameter. This paradigm is not, however, founded on a compelling
underlying theory and therefore it is worthwhile exploring other 
possibilities.

In this paper we focus on 
Anomaly Mediation (AM)~\cite{Randall:1998uk}-\cite{Pomarol:1999ie}. 
This is a framework in 
which a single mass parameter determines the $\phi^*\phi$, $\phi^3$ and
$\lambda\lambda$ \sy-breaking terms in terms of calculable and, 
moreover, renormalisation group (RG) invariant functions of the 
dimensionless couplings, in an elegant and predictive way; too
predictive, in fact, in that the theory in its simplest form leads to
tachyonic sleptons and fails to accommodate the usual
electroweak vacuum state. There is a natural solution to this, however,
which restores the correct vacuum while retaining the RG invariance
(and hence the ultra-violet insensitivity) of the predictions. This is
achieved simply (and without introducing another source of explicit 
\sy-breaking) by the introduction of an additional anomaly-free gauged $U(1)'$, 
broken at a high scale, so that the mass contributions from the
$U(1)'$ $D$-term are naturally of the same order of magnitude as the 
soft breaking terms, and can eliminate the tachyonic slepton problem.
This scenario was first explored in any detail in
Ref.~\cite{Jack:2000cd}, and subsequently by a number of 
authors~\cite{Arkani-Hamed:2000xj}-\cite{Jones:2006re}.

In~\reference{Hodgson:2005en}\ it was shown how the characteristic low energy
theory can arise in a natural way from a $U(1)'$ with a Fayet-Iliopoulos
(FI) $D$-term. The scale of the $U(1)'$ breaking was associated with 
the scale of the FI term, but, as indicated above, the resulting 
mass contributions to the \mssm\ scalars are naturally of order the \sy-breaking scale. 

In \reference{Jones:2006re}, it was shown how it was possible to 
work with a superpotential which was more complicated but still 
purely cubic, and nevertheless dispense with the FI term. The $U(1)'$ breaking scale 
was then generated by dimensional transmutation in a way reminiscent of the geometric hierarchy 
model of Witten~\cite{Witten:1981kv}. 

Here we return to an improved version of the original model of
\reference{Hodgson:2005en}, augmented by the introduction in the
superpotential of a linear term for the gauge singlet field $S$. We will
see that this incarnation of the model is natural, in the sense that the
superpotential contains  every possible term allowed both by the $U(1)'$
and a $U(1)_R$ global symmetry. The coefficient  of the $S$ linear term
provides an alternative to the FI term for setting the scale  of the
$U(1)'$ breaking. 

In this paper we explore some cosmological consequences which flow
naturally from the introduction of $U(1)'$. We will see that the
minimal acceptable form of the theory incorporates a natural mechanism
for supersymmetric $F$-term inflation, although with a potential
somewhat more complicated than the simple models in the literature. 
Inflation ends with a phase transition producing gauge cosmic strings,
and hence tight constraints on the parameters from the Cosmic Microwave
Background \cite{Battye:2010hg}. In common with other \sy{} models with
conserved $R$-parity, it naturally accommodates a Dark Matter candidate,
which is mostly wino. Although it has been argued that in \amsb{} models
the annihilation cross-section is too large for conventional freeze-out
to generate the dark matter, there is another source of dark matter in
the form of particles radiated by cosmic strings. The model also has the
possibility of baryogenesis via leptogenesis or the Affleck-Dine
mechanism, with \cp\ violation supplied by the neutrino sector. 
Gravitinos are very massive, and so decay early enough not to be in
conflict with nucleosynthesis.

The model has the same field content as the $F_D$ hybrid inflation
model~\cite{Garbrecht:2006,Garbrecht:2006az}, 
but different charge assignments and
couplings. $F_D$ hybrid inflation also has a singlet which is a natural
inflaton candidate, but differs in other ways: for example, right-handed
neutrinos have electroweak-scale masses, and the gravitino problem is
countered by entropy generation.

Since we entertain the possibility of a Fayet-Iliopoulos term in our
model, it behoves us to address issues raised by the recent papers on
the connection of FI terms with supergravity, by Komargodski and
Seiberg, and others~\cite{Komargodski:2009pc}-\cite{Distler:2010zg}. The
upshot of this work is the conclusion that a global theory with a FI
term cannot be consistently embedded in supergravity. An exception
would be allowed  if the full supergravity theory has a certain exact
continuous global symmetry; but no consistent theory of gravity is
allowed to have such a symmetry. However, this latter assertion (that
global symmetries are forbidden from theories with gravity) has the
status of a (admittedly widely believed) conjecture rather than a proven
theorem \footnote{We thank Tom Banks for conversations on this topic.}.
For a relevant and detailed analysis see \reference{Banks:2010zn}. We
propose to exploit this loophole to justify this aspect of 
our discussion, and argue that
our model has enough interesting features to render this exercise 
worthwhile. One such feature is that, as we shall see, our model indeed
has an exact global $U(1)_R$ symmetry; although it is not clear that
this symmetry (or indeed the global version of $U(1)'$) is of the type
required by \reference{Komargodski:2009pc}. We will, however, also see
that our model is perfectly viable if in fact it has no FI term.

\section{The \amsb{} soft terms}

We will assume that supersymmetry breaking arises via the renormalisation 
group invariant form characteristic of Anomaly Mediation, 
so that the soft parameters for the gaugino mass $M$, the $\phi^3$ interaction 
$h$ and the $\phi^*\phi$ and $\phi^2$ mass terms $m^2$ and $m_3^2$ 
take the generic form
\bea
M_i & = & m_{\frac{3}{2}} \beta_{g_i}/{g_i}\label{eq:AD1}\\
h_{U,D,E,N} & = & -m_{\frac{3}{2}}\beta_{Y_{U,D,E,N}}\label{eq:AD2}\\
(m^2)^i{}_j &=& \frac{1}{2}m_{\frac{3}{2}}^2\mu\frac{d}{d\mu}\gamma^i{}_j
+kY_i\delta^i{}_j,\label{eq:AD3}\\ 
m_3^2 & = & \kappa m_{\frac{3}{2}} \mu_h - m_{\frac{3}{2}} \beta_{\mu_h}.
\label{eq:AD4}
\eea
Here $\mu$ is the renormalisation scale, 
$m_{\frac{3}{2}}$ is the gravitino mass, which 
to obtain a reasonable \sic\ spectrum 
will be typically $40-50\TeV$. 
$\beta_{g_i}$ are the gauge $\beta$-functions and $\gamma$ the 
chiral supermultiplet anomalous dimension matrix. 
$Y_{U,D,E,N}$ are the $3 \times 3$ Yukawa matrices,
while the $Y_i$ are charges corresponding to a $U(1)$ symmetry of the
theory with no mixed anomalies with the gauge group; the $kY$ term
corresponds in form to a FI $D$-term. As we shall see, we can generate
this $kY$ term naturally with $k$ of $O(m_\frac{3}{2}^2)$, by breaking
a $U(1)'$ symmetry at a large scale. The parameter $\kappa$ is an
arbitrary constant; its presence means that in sparticle spectrum
calculations one is free to determine $m_3^2$ (and the value of the
Higgs $\mu$-term, $\mu_h$) by minimising the Higgs potential at the electroweak
scale in the usual way.

\section{The $U(1)'$ symmetry}
The \mssm{} (including right-handed neutrinos) admits 
two independent generation-blind
anomaly-free $U(1)$ symmetries\footnote{One of the attractive features of minimal 
Anomaly Mediation is that squark/slepton mediated flavour changing neutral 
currents are naturally small~\cite{Allanach:2009ne}; this feature is preserved 
by a generation-blind $U(1)$ but not by a flavour-dependent $U(1)$, so 
we stick to the former here.}. 
The possible charge assignments are shown in Table~1.
\begin{table}
\begin{center}
\begin{tabular}{|c|c c c c c c|} \hline
&$Q$ & $U$ & $D$
& $H_1$ & $H_2$ & $N$ \\ \hline
&& & & & & \\ 
$q$&$-\frac{1}{3}q_L$ & $-q_E-\frac{2}{3}q_L$ & $q_E+\frac{4}{3}q_L$
& $-q_E-q_L$ & $q_E+q_L$ & $-2q_L-q_E$ \\ 
&& & & & & \\ \hline
\end{tabular}
\caption{\label{anomfree}Anomaly free $U(1)$ symmetry for arbitrary
lepton doublet and singlet charges $q_L$ and $q_E$ respectively.}
\end{center}
\end{table} 
The \sm{} gauged $U(1)_Y$ is $q_L=1, q_E = -2$; this $U(1)$ is of
course anomaly free even
in the absence of $N$. $U(1)^{B-L}$ is
$q_E = -q_L = 1$; in the absence of $N$ this would have $U(1)^3$ and
$U(1)$-gravitational anomalies, but no mixed anomalies with the \sm{} gauge
group. 

Our model will have, in addition, a pair of \mssm{} singlet fields 
$\Phi, \Phib$ with $U(1)'$ charges
$q_{\phi, \phib} = \pm (4q_L+2q_E)$ and a gauge singlet $S$.
In order to solve the tachyon slepton problem we will need 
that, for our new gauge symmetry $U(1)'$, the charges $q_L,q_E$ have the 
same sign; where numerical results are required
we will use $q_L=2, q_E=1$. 
The resulting hypercharges are shown in Table 2. 
\begin{table}
\begin{center}
\begin{tabular}{|c| c c c c c c c c c c c|} \hline
$U(1)$ &$Q$ & $U$ & $D $ & $L $& $E$
& $H_1$ & $H_2$ & $N$& $\Phi $& $\Phib$& $S $ \\ \hline
$Y$ &1/3 & -4/3 & 2/3 & -1 & 2
& -1 & 1 & 0& 0& 0& 0
 \\ \hline
$Y'$ &-2/3 & -7/3 & 11/3 & 2& 1
& -3 & 3 & -5& 10&-10& 0 \\ \hline
\end{tabular}
\caption{\label{anomfreeb}
Hypercharges for $U(1)_Y$ and $U(1)'$.}
\end{center}
\end{table}
Finally, note that with {\it any}\ charge assignment such that $q_{L,E} > 0$, 
only $E,L,D,H_2,\Phi$ have {\it positive}\
$U(1)'$ charges. This will be important in what follows when we come to
consider the inflationary regime.

\section{The superpotential and spontaneous $U(1)'$ breaking}

The complete superpotential for our model is:
\ben 
W = W_A + W_B
\label{eq:superpot}
\ee
where $W_A$ is the \mssm\ superpotential, omitting the Higgs 
$\mu$-term, and augmented by 
Yukawa couplings for the right-handed neutrinos:
\ben
W_A = H_2 Q Y_U U + H_1 Q Y_D D + 
H_1 L Y_E E + H_2 L Y_{N} N 
\ee
and
\ben
W_B = \lambda_1 \Phi\Phib S + \half\lambda_2 N N \Phi 
+ \lambda_3 S H_1 H_2 +M^2 S,
\ee
where $M,\la_1,\la_3$ are real and positive 
and $\lambda_2$ is a symmetric $3\times 3$ matrix.
This model differs
from the one first considered in Ref.~\cite{Hodgson:2005en} by the inclusion of
the $\lambda_3$ term and the linear term for $S$. Although the old
model led to satisfactory $U(1)'$ breaking, it had several defects from
a cosmological perspective; in particular a second stable particle in additional to
the usual \lsp, because its superpotential was invariant with respect to
\ben 
\Phib \to - \Phib, S \to -S
\ee
with all other fields unchanged. Now in the \amsb\ context, an
additional Dark Matter candidate would not be a bad thing per se, since
(in \amsb) the \lsp\ is generally the neutral wino, whose annihilation
cross-section is generally considered to be too high for comfort.
However, the additional stable particle in the original formulation of
the model has a mass at the $U(1)'$-breaking scale, which, as we will
see, we are going to want to be rather large. 

Note that the $U(1)'$ symmetry forbids the renormalisable $B$ and $L$
violating superpotential interaction terms of the form $QLD$, $UDD$,
$LLE$, $H_1 H_2 N$ and $N S^2$, $N^2 S$ and $N^3$, 
as well as the mass terms $N S$, $N^2$ and $L
H_2$ and the linear term $N$. Moreover $W_B$ contains the only 
cubic term involving $\Phi,\Phib$ that is allowed. 
Our new superpotential \eqn{eq:superpot}\
in fact is completely natural, in the sense that it is invariant under
a global $R$-symmetry, with superfield charges 
\ben
S=2, L=E=N=U=D=Q=1, H_1= H_2 = \Phi = \Phib = 0,
\label{rsym}
\ee
which forbids the remaining gauge invariant 
renormalisable terms ($S^2$, $S^3$, $\Phi\Phib$ and $H_1 H_2$). 
Moreover, this $R$-symmetry forbids the quartic superpotential
terms $QQQL$ and $UUDE$, which are allowed by the $U(1)'$ symmetry,
and give rise to dimension 5 operators capable of
causing proton decay~\cite{Weinberg:1981wj}-\cite{Sakai:1981pk}. 
It is easy to see, 
in fact, that the charges in \eqn{rsym}\ 
disallow $B$-violating operators in the superpotential 
of arbitrary dimension. 
Of course this $R$-symmetry is broken by the soft \sy\ breaking.

Retaining for the moment only the scalar fields $\phi, \phib, s$ 
(the scalar component of their upper case counterpart 
superfields) we write the scalar potential:
\bea
V &=& \lambda_1^2 (|\phi s|^2 + |\phib s|^2) + |\lambda_1\phi\phib+M^2|^2
+\frac{g'^2}{2}\left(\xi - q_{\phi}|\phi|^2 + q_{\phi}|\phib|^2\right)^2\nn
&+&m_{\phi}^2 |\phi|^2 + m_{\phib}^2 |\phib|^2 +m_s^2|s|^2 
+\rho M^2 m_{\frac{3}{2}} (s + s^*)\nn
&+& h_{\lambda_1}\phi\phib s + c.c..
\eea
Here the $h_{\lambda_1}$ term is a soft breaking, determined in accordance with 
\eqn{eq:AD2}:
\ben
h_{\lambda_1} = - m_{\frac{3}{2}}\frac{\la_1}{16\pi^2}\left(
3\la_1^2 +\frac{1}{2}\Tr \lambda_2^2 +2\lambda_3^2 - 4q_\phi^2{g'}^2 \right),
\een
denoting the $U(1)'$ charge by $g'$. 
We also introduce a soft breaking 
term linear in $s$ (see~\cite{Jack:2001ew} for a discussion of linear 
terms in this context), and a Fayet-Iliopoulos term $\xi$ for $U(1)'$.
Our model thus has two explicit mass parameters $M, \sqrt{\xi}$ as well
as the gravitino mass $m_{\frac{3}{2}}$. It is easy to show that if 
$M \gg m_{\frac{3}{2}}$, then $V$ above is minimised with
$\phi, \phib$ acquiring large vevs, while $s$ acquires a vev of 
$O(m_{\frac{3}{2}})$. Thus the right-handed neutrinos acquire large
masses, and an appropriate $\mu$-term for the Higgs doublets is
generated in the manner of the \nmssm{} (for a review of and references for 
the \nmssm{} see Ref.~\cite{Ellwanger:2009dp}).

At the 
minimum of $V$ we have
\bea
\lambda_1 v_{\phi}v_{\phib} +M^2 &\approx& 0,\label{vmina}\\
\xi - q_{\phi}v_{\phi}^2 + q_{\phi}v_{\phib}^2 &\approx& 0,\label{vminb}\\
v_s &\approx& -\frac{h_{\lambda_1} v_{\phi}v_{\phib} 
+ \rho M^2 m_{\frac{3}{2}}}{v_{\phi}^2 + v_{\phib}^2}
\label{vminc}
\eea
and we see that $v_s$ is $O(\la_1m_{\frac{3}{2}}/(16\pi^2))$ 
if $M \gg \sqrt{\la_1\xi}$ or if $M \sim \sqrt{\la_1\xi}$, 
and $v_s$ is $O(m_{\frac{3}{2}} M^2/(16\pi^2\xi))$ if $M \ll \sqrt{\la_1\xi}$.
(Here we assume that $\rho$, like $h_{\lambda_1}$, 
is suppressed by a loop factor, and that $q_\phi g' \sim O(1)$). 
For large $M$ and/or $\xi$, (all trace of the
$U(1)'$ in the effective low energy Lagrangian disappears, except for
contributions to the masses of the matter fields, arising from the
$U(1)'$ $D$-term, which are naturally of the same order as the \amsb{} 
ones. If we neglect terms of $O(m_{\frac{3}{2}})$, the breaking of
$U(1)'$ preserves supersymmetry; thus the $U(1)'$ gauge boson, its 
gaugino (with one combination of $\psi_{\phi,\phib}$) and the Higgs
boson form a massive supermultiplet with mass 
$m \sim g'\sqrt{v_{\phi}^2 + v_{\phib}^2}$,
while the remaining combination of $\phi$ and $\phib$ 
and the other combination of $\psi_{\phi,\phib}$ form a massive
chiral supermultiplet, with mass 
$m \sim \lambda_1\sqrt{v_{\phi}^2 + v_{\phib}^2}$.

Evidently $s$ also gets a large \sic\ mass, as does the $N$ triplet, thus
naturally implementing the see-saw mechanism. Moreover, the vev for $s$
introduces a Higgs $\mu$ term, 
the magnitude of which is naturally related to the \sy-breaking scale.

It is easy to show that the contribution to the slepton masses 
arising from the $U(1)'$ term which resolves the tachyonic slepton problem
is given (after spontaneous breaking of the $U(1)'$) by~\cite{Hodgson:2005en} 
\ben
\delta m^2 \sim -\frac{q_{L,e}}{q_{\phi}}m_{\phi}^2,
\label{eq:tachs}
\een
and also (using Eqs.~(8), (10) and Fig.~1 of 
\reference{Hodgson:2005en}) that we need 
\ben
\delta m^2 \sim
0.1 \left(\frac{m_{\frac{3}{2}}}{40\TeV}\right)^2.
\een 
So, if we assume that the one-loop $\gamma_{\phi}$ is dominated 
by its gauge contribution, then we have from 
\eqn{eq:AD3}\ that 
\ben
m_{\phi}^2 \sim m_{\frac{3}{2}}^2 \beta_g' \frac{\pa}{\pa g'}\gamma_{\phi} 
\sim 
- m_{\frac{3}{2}}^2 Q q_{\phi}^2 g'^4/(16\pi^2)^2
\een
where here $Q$ is the sum of the squares of all the $U(1)'$ charges. 
So we would want 
\ben
q_{L,e} Q q_{\phi} g'^4\sim 
\left(\frac{m_{\frac{3}{2}}}{40\TeV}\right)^2
\label{eq:gprimevalue}
\een
at the \sy\ breaking scale. 

Let us use $q_L = 2 ,q_E = 1$, as mentioned previously. Then, 
$Q = 516$, and it is easy to show using the RG equation for 
$\beta_{g'}$ that if $g' = 1$ at $10^{16}\GeV$, then 
at $10^3\GeV$, $g' \sim 0.1$ and that consequently 
\eqn{eq:gprimevalue} is reasonably well satisfied (for 
$m_{\frac{3}{2}}= 40\TeV$). 
Subsequently we will use $g' = 1$ at high energies. 

It is quite interesting to contrast our model with the conventional 
versions of the \nmssm, which does not, in its basic form, contain an
extra $U(1)$, but where a vev (of the scale of \sy\ breaking) for the gauge
singlet $s$ generates a Higgs $\mu$-term in much the same way, as is
done here. However, while in the \nmssm\ case the $s$ fields are very
much part of the Higgs spectrum, here, in spite of the
comparitively small $s$-vev, the $s$-quanta obtain large \sic\ masses and
are decoupled from the low energy physics, which becomes simply that
of the \mssm. Another nice feature is the natural emergence of the seesaw 
mechanism via the spontaneous breaking of the $U(1)'$. These features of the 
model still hold for $\xi = 0$, as 
can easily be seen from Eqs.~(\ref{vmina})-(\ref{vminc}). 
Moreover, the FI-type mass contributions 
of the form of \eqn{eq:tachs}\ are still present. This makes the model 
an interesting one from the particle physics point of view both with and without 
the FI term. In the latter case we of course have no problem with 
the conclusions of \reference{Komargodski:2009pc}; but we also discuss the former case also because of the 
novel nature of the inflationary regime in that case.

\section{{The scalar potential at large $s$.}\label{effpot}}

As we have developed it, the theory naturally provides for $F$-term 
inflation~\cite{Copeland:1994vg}-\cite{Lyth:1998xn}. Although we have a
$U(1)'$, there is a crucial difference between our scenario and the
original $D$-term inflation paradigm~\cite{Binetruy:1996xj}, which is
that the \sm{} fields are necessarily charged under $U(1)'$. 
Consequently, it is possible for the large $s$ tree potential to be
$D$-flat for the whole gauge group, including $U(1)'$; even in the
presence of a Fayet-Iliopoulos term, which is one thing that makes
this case of interest.

The tree potential is 
(each term below involves implicit summation
over all indices, including generation, not involved in explicit manipulations, 
and $d^c$, for example, represents all three generations 
of $SU_2$ singlet (anti-)down squarks):

\bea
V_{\rm tree} &=& \lambda_1^2 |\phi s|^2 + |\lambda_1 \phib s 
+\frak{1}{2}\lambda_2 \nu^c\nu^c|^2 \nn
&+& |\lambda_2 \nu^c \phi + lY_{N}h_2|^2 + 
|\lambda_3 h_1 h_2 + \lambda_1\phi\phib+M^2|^2\nn
&+&|\lambda_3 h_1 s +q Y_U u^c + l Y_{N}\nu^c |^2 
+|\lambda_3 h_2 s +q Y_D d^c + l Y_E e^c |^2\nn 
&+&|Y_U u^c h_2 + Y_D d^c h_1|^2 
+|Y_E e^c h_1 + Y_{N} \nu^c h_2|^2\nn 
&+&\frac{1}{2}g'^2\bigg(\xi - q_{\Phi}\phi^*\phi - q_{\Phib}\phib^*\phib
 - q_{Q} q^{\dagger}q - q_{D} d^{c\dagger}d^c - q_{U} u^{c\dagger}u^c\nn
&-& q_{L} l^{\dagger}l- q_{E} e^{c\dagger}e^c- q_{N} \nu^{c\dagger}\nu^c-
q_{H_1} h_1^{\dagger}h_1- q_{H_2} h_2^{\dagger}h_2\bigg)^2\nn
&+& \frak{1}{8}g_3^2 \sum_{a}\left(q^{{\dagger}}\lambda^a q
+d^{c{\dagger}}(-\lambda^a)^T d^c +u^{{c\dagger}}(-\lambda^a)^T u^c \right)^2 \nn
&+& \frak{1}{8}g_2^2 \sum_{a}(q^{\dagger}\sigma^a q+l^{\dagger}\sigma^a l +h_1^{\dagger}\sigma^a h_1 
+h_2^{\dagger}\sigma^a h_2 )^2 \nn
&+& \frak{1}{8}g_1^2 (\frak{1}{3}q^{\dagger}q-\frak{4}{3}u^{c\dagger}u^c
+\frak{2}{3}d^{c\dagger}d^c-l^{\dagger}l+2e^{c\dagger}e^c - h_1^{\dagger}h_1
 + h_2^{\dagger}h_2)^2\nn
&+& V_\mathrm{soft}.
\label{eq:treepot}
\eea
Here $V_\mathrm{soft}$ contains the \amsb{} soft terms, which are suppressed 
by at least one power of $m_{\frac{3}{2}}$, that is 
\ben
V_\mathrm{soft}= \rho M^2 m_{\frac{3}{2}} (s + s^*)
+m_s^2 |s|^2+m_{\phi}^2 |\phi|^2 + m_{\phib}^2 |\phib|^2
+m_{l}^2 |l|^2 + \cdots.
\ee
Note that in \eqn{eq:treepot}\ we have written the $U(1)_Y$ gauge coupling 
as $g_1$, although its normalisation corresponds to the usual \sm\ convention, not 
that appropriate for $SU(5)$ unification. This is to avoid confusion with the 
$U(1)'$ coupling, $g'$.

At large fixed $s$ we see that there are mass terms proportional to
$\lambda_1^2 |s|^2$ for $\phi, \phib$ which will mean that in this region 
their vevs will be zero. (Actually this is a rather more subtle point 
than it might appear; we will discuss it in more detail in
section~(\ref{fieldspace}).) In the presence of a $U(1)'$ FI term, 
we introduce $s$-independent vevs for some of the \mssm{} scalars in such a
manner as to achieve $D$-flatness for $SU(3)_c\otimes SU(2)_L \otimes
U(1)_Y\otimes U(1)'$. 

\subsection{Where in field space is the minimum?}\label{fieldspace}

The minimum of our tree potential is
in fact very degenerate, like the well known case the \mssm. Let us
begin by enumerating the degrees of freedom represented in the
potential. A naive count would suggest that there are 55 complex
$F$-terms; one for each complex degree of freedom (c.d.o.f.)). In addition 
there are 13 real $D$-terms, and 13 (real) gauge choices, one of each for
each generator of the symmetry. This should (naively) lift all
degeneracy. However, as in the \mssm, choosing $\langle
H_{1,2}^\alpha\rangle=0$ kills the $F$-terms for all \sm{} fields except the
Higgs fields themselves, and thus the degeneracy arises. Now we have 51
c.d.o.f. with only 10 $F$-terms left, and 13 $D$-terms i.e.\ a 28 complex
dimensional degenerate space.

Let us look at the cases of large $s$ (inflation) and
small $s$ (today) separately. The latter case was addressed in
section 4, but it is possible that there are other minima. 
For example, $F_S$ could be made zero by giving the Higgs fields
large vevs instead of $\phi,\phib$. That would require the remaining
\mssm{} scalars to have zero vevs. Alternatively, if the Higgs fields
had zero vevs, one could have a vev in some combination of \mssm{} scalar
fields, just as in the scenario we in fact pursue 
for inflation. We shall not investigate this
further; we have a good local minimum 
in hand, which gives us the \sm{} physics we know and an explanation for
other things as well, like the Higgs $\mu$-term, and neutrino masses and mixings. 
Whether our minimum is favoured after loop corrections, we shall
not investigate.
But it is not clear that we are in the global minimum anyway.

For $M,\sqrt{\xi} \ll \langle s\rangle$ i.e. during inflation, we shall
focus on a field space region with $\vev\phi = \vev\phib = 0$, so that
$V_{\rm tree} = M^4$, with an appropriate set of \mssm\ field vevs
arranged to render all the $D$-terms (including the $U(1)'$ term, 
which for $\xi \neq 0$ must have some non-zero vevs) and $F$-terms
(excluding $F_S$) zero. This is naturally motivated by the presence in
$V$ of $|\phi s|^2$ and $|s\phib|^2$ terms. In fact, however, it does
not represent an absolute minimum of $V_{\rm tree}$; it is easy to show
that by choosing $\vev\phib\sim s^N$, $\vev\phi\sim s^{-N}$ and $\nu^c
\sim s^{\frac{N+1}{2}}$ for some $N > 1$, with other vevs chosen so as
to achieve $D$-flatness, it is possible to make $V_{\rm tree}$
arbitrarily small for $s$ large. Nevertheless, $\vev\phi = \vev\phib =
0$ does represent a local minimum of the tree potential, for $s^2 >
s_c^2$ where 
\ben
 s^2_c = M^2/\la_1.
\label{scrit}
\een
As we shall see, this value for $s$ corresponds
to the appearance of a zero eigenvalue in the $\phi,\phib$ mass system, and we believe the
radiative corrections lead to decrease towards this critical
point. (Certainly the afore-mentioned $|s\phib|^2$ term
in $V$ will discourage evolution towards large $\phib$).

The flat space (the large space with $V=V_{min}$) in the \mssm{} is
complicated because it is not additive in field space - ie. the sum of
two position vectors each pointing to a place in the flat space is not
(necessarily) a position vector pointing to a place in the flat space. 
But it is, of course, scalar multiplicative (in the \mssm), ie. a position vector
pointing to a place in the flat space multiplied by any scalar is still
a position vector pointing to a place in the flat space. In our model,
the flat space loses this virtue when $\xi \neq 0$; then even the
origin in field space is not part of the flat space, since it has
$D'=g'\xi$. In any event, it is hopeless to parameterise a 28 complex
dimensional space. In contrast to the \mssm{}, which has no mass scales
(except the Higgs $\mu$-term), our model has $\langle s
\rangle,M,\sqrt{\xi}$ which contribute mass terms to scalars, fermions
and vectors. Moreover, since $F_S \neq 0$, loop-corrections can lift the
degeneracy, in contrast to in the \mssm.

It is not feasible to parameterise this large space, but we can get a
taste of it, by showing a 4 dimensional subspace:
\bea
\vev{u_2} = \vev{c_3} = \vev{t_1}
&=& \Delta_A\sqrt{\frac{\xi}{5}},\nn
\vev{d_3} = \vev{s_1} = \vev{b_2}
&=& \Delta_B\sqrt{\frac{\xi}{5}},\nn
\vev{u^c_1} = \vev{c^c_2} = \vev{t^c_3}
&=& \Delta_C\sqrt{\frac{\xi}{5}},\nn
\vev{d^c_1} = \vev{s^c_2} = \vev{b^c_3}
&=& \Delta_D\sqrt{\frac{\xi}{5}},\nn
\vev{\nu_e} &=& \sqrt{1+2\Delta_A^2-\Delta_B^2+\Delta_C^2
-2\Delta_D^2}\sqrt{\frac{\xi}{5}},\nn
\vev{\mu}&=&
\sqrt{1-\Delta_A^2+2\Delta_B^2
+\Delta_C^2-2\Delta_D^2}\sqrt{\frac{\xi}{5}},\nn 
\vev{\tau^c} &=& \sqrt{1
+3\Delta_C^2-3\Delta_D^2}\sqrt{\frac{\xi}{5}}. \label{eq:FvevsLARGE}
\eea 
We have now changed the notation for the fields so as to distinguish the generations 
by name, and explicitly indicate the colour index on the down squarks.
Here we see we have given vevs to all superfields 
that can have one: $Q,U,D,L,E$ 
We recognise $\Delta_A=\Delta_B=\Delta_C=\Delta_D=0$ as $LLE$ and
$\Delta_A=\Delta_B=\Delta_C=0, \Delta_D=1/\sqrt{3}$ as $DDDLL$. 
We can also see that \sm{} flat directions $QQQL,UUUEE$ are
present, but the parameters cannot make either of these alone. That is
because they have the wrong sign of $U(1)'$ charge, and thus cannot
balance $\xi$. 

Even this system is too complicated to analyse in general; one could
choose some cases to try for specific values of the 4 parameters, but
we shall not pursue this further here. Rather, we shall look at a
specific subspace of the flat space with just 1 free parameter, and
investigate how the the loop potential depends on it. 

We put vevs in $DDDLL$ and $LLE$ invariants, by having nonzero vevs only 
for $L_{1,2},E_3,D_{1,2,3}$, where the label denotes generation. 
Specifically, we set
\bea \vev{\tau^c} = v_E &=& \Delta\sqrt{\frac{\xi}{15}},\nn
\vev{\nu_e} = \vev{\mu} = v_L &=& \sqrt{(1 + \frac23
\Delta^2)\frac{\xi}{15}},\nn 
\vev{d^c_1} = \vev{s^c_2} =
\vev{b^c_3} = v_D &=& \sqrt{(1 - \frac13 \Delta^2)\frac{\xi}{15}}
\label{eq:Fvevs} 
\eea 
where we introduce the more convenient parameter $\Delta=\sqrt{3-9\Delta_D^2}$
which is real and satisfies $0\leq \Delta \leq \sqrt{3}$.

By working in a ``basis'' such that $Y_{U,D,E,N}$ are diagonal, we ensure that,
(given \eqn{eq:Fvevs}) 
there is no contribution to $V_{\rm tree}$ from any $F$-term except $F_S$, and 
the \ckm-matrix is the identity matrix. 
This is not an approximation as such, one can adjust for the influence of the mixings
by choosing vevs that are rotated correspondingly. 
Note that the vevs are in ``colour=generation'' for $D$ and ``weak
charge=generation'' for $L$, and in the third generation for $E$. Since each
superfield only has one nonzero entry, flatness is independent of 
the phases of the fields.
One can use the gauge choices of the diagonal generators to remove one
phase each from the vevs of the fields. The 5 gauge choices are made so as to remove
all phase differences in the vevs.
To make the result even simpler, we have chosen the common phase to be zero; all vevs
real and positive (this corresponds to a choice for the global U(1)
symmetry of the \sm).
We have also taken the vev of S to be real - this defines the
coordinate system for the couplings of S.

The combination of zero vevs for $\phi,\phib$ with the set of vevs described in 
\eqn{eq:Fvevs}\ means that for large $s$ we have (neglecting the tree soft terms)
\ben
V_{\rm tree} = M^4.
\een
With the inclusion of one loop corrections this becomes 
\ben
V = M^4 + \Delta V,
\ee
where $\Delta V$ represents the one-loop corrections, given as usual by 
\ben
\Delta V = \frac{1}{64 \pi^2} 
{\rm Str} (M^2(s))^2\ln(M^2(s)/\mu^2).
\label{Vone}
\ee
Here 
\ben
{\rm Str} \equiv \sum_{\rm scalars} - 2 \sum_{\rm fermions} 
+ 3 \sum_{\rm vectors}.
\ee
Contributions to $\Delta V$ from fields with large masses will be more 
significant than those from the neglected soft terms; but 
of course for fields which, although massive, 
form degenerate supermultiplets, the contributions to $\Delta V$ will cancel 
exactly. It is easy to see that there are two relevant sets of contributions.

\subsection{The $\Phi,\Phib$ system}\label{phivevs}

Let us consider the $\Phi$, $\Phib$ subsystem,
which in fact appears in the minimal $F$-term inflation model\cite{Copeland:1994vg}.
The scalar mass matrix eigenvalues at large fixed $s$
are given by 
\ben M_{\phi,\phib}^2 = \la_1^2s^2 \pm
\lambda_1 M^2, \quad \hbox {(twice each)}. 
\label{phimasses}
\een 
(Note that there is no contribution to these mass
terms from the $U(1)'$ $D$-term, because of the $D$-flatness
engendered by \eqn{eq:Fvevs}. But of course as a result of these vevs 
there will be further significant contributions to the one-loop potential beyond those 
considered in this subsection; these we will describe in the next one). 
The corresponding fermion masses are
simply $\mtilde^2_{\tilde{\phi},\tilde{\phib}} = \lambda_1^2 s^2.$ 
The contribution to the one-loop scalar potential is 
\bea
\Delta V_1 
&=&\frac{1}{32\pi^2}\bigl[(\lambda_1^2 s^2+\lambda_1 M^2)^2\ln 
\left(\frac{\la_1^2s^2+\lambda_1 M^2}{\mu^2}\right)
+ (\lambda_1^2 s^2-\lambda_1 M^2)^2\ln 
\left(\frac{\la_1^2s^2-\lambda_1 M^2}{\mu^2}\right)\nn 
&-&2\lambda_1^4 s^4\ln\left(\frac{\la_1^2s^2}{\mu^2}\right)
\bigr].
\eea
If we assume that
we are interested in values of $s$ for which $\la_1 s^2 \gg M^2$, 
it is easy to show that this reduces to 
\ben \Delta V_1 = 
\frac{1}{16\pi^2}\lambda_1^2 M^4\ln \left(\frac{\la_1^2s^2}{\mu^2}\right). 
\label{Vphiphib}
\een

\subsection{The $(H_1,H_2,Q,E_{1,2},L_3,N_{1,2})$ system}\label{sec:quatics}

Apart from the $\Phi, \Phib$ system already considered, the only other 
contributions to the one loop potential comes from the 
$(H_1,H_2,Q,E_{1,2},L_3,N_{1,2})$ system, where the scalar mass matrix 
can be split into two separate $12\times 12$ complex matrices. 
Note that it is one particular linear combination of the three 
doublets $Q$ which is selected by the $D$-vevs; thus if we define 
\ben
Q = \frac{y_d Q_1 + y_s Q_2 + y_b Q_3}{\sqrt{y_d^2 +y_s^2 + y_b^2}}
\ee
then the Higgs-squark doublet mixing term is 
\ben
\lambda_3 y s v_D q^{\dagger} h_2 + c.c., 
\ee
where $y = \sqrt{y_d^2 +y_s^2 + y_b^2}$. The
first scalar matrix ($h_1^1,h_2^2,q,
\mu^c,\tau,\nu_e^c$) takes the form

\begin{eqnarray}\label{bigmat}
\begin{pmatrix}
M_S^2+M_D^2+M_{E_2}^2+M_{E_3}^2	& M_M^2\cdot\si_1	& \cdot	& \cdot	& \cdot &M_{\nu_1} M_S	\\
M_M^2\cdot\si_1	& M_S^2+M_{\nu_1}^2	&M_S M_D		& M_S M_{E_2}& M_S M_{E_3}	& \cdot	\\
\cdot 		& M_S M_D 		&M_D^2	& M_D M_{E_2}	&M_D M_{E_3} & \cdot	\\
\cdot 		& M_S M_{E_2} 		& M_D M_{E_2}	& M_{E_2}^2	&M_{E_2} M_{E_3}& \cdot	\\
\cdot 		& M_S M_{E_3} 		& M_D M_{E_3}	& M_{E_2} M_{E_3}	&M_{E_3}^2& \cdot	\\
M_{\nu_1} M_S			& \cdot		& \cdot	& \cdot	&\cdot& M_{\nu_1}^2 

\end{pmatrix} \; 
\end{eqnarray}
where 
\ben
M_S = \lambda_3 s, M_D = y v_D, M_{E_2} = y_{\mu} v_L, M_M^2 = M^2\lambda_3, 
M_{\nu_1} = y_{\nu_e} v_L, M_{E_3}=y_{\tau} v_E.
\ee
$\si_1$ is the usual Pauli matrix, and if no $2\times 2$ matrix is
indicated the identity matrix is to be assumed. A dot indicates the zero
matrix. The second matrix is identical except that $M_M^2$ is replaced
by -$M_M^2$, $y_{\nu_e}v_L$ is replaced by $y_{\nu_{\mu}}v_L$, and
$y_{\mu}v_L$ is replaced by $y_e v_L$. 

The eigenvalue equation for the matrix \eqn{bigmat}\ has four zero
eigenvalues; the rest of it can be factorised into a product of two identical
quartic equations of the form
\ben
x^4 - 2a_3 x^3 + (a_3^2 + 2a_2 a_1 - a_0^2)x^2 
-(2a_3 a_2 a_1 - (a_2+a_1)a_0^2)x + a_2 a_1 (a_2 a_1 - a_0^2) = 0,
\label{quartic}
\ee
where
\bea
a_0 &=& M_M^2,\nn
a_1 &=& M_{E_2}^2+M_D^2+M_{E_3}^2,\nn
a_2 &=& M_{\nu_1}^2,\nn
a_3 &=& M_S^2+a_2+a_1.
\label{qcoeffs}
\eea 
For a quartic equation 
with non-zero real coefficients, of the form
$x^4-ax^3 + bx^2 -cx +d=0$, and all roots known to be real, the necessary and 
sufficient conditions
that all its roots be positive are $a,b,c,d > 0$
\footnote{Corollary of Descartes' rule of signs.}. We see from 
\eqn{quartic}\ that these conditions are satisfied provided
\ben
a_2 a_1 > a_0^2 .
\label{cond}
\een
Inserting the vevs, we require
\ben
\left(y^2 \left(1-\frac{1}{3}\Delta^2\right)+y_{\mu}^2 \left(1+\frac{2}{3}\Delta^2\right)
+y_{\tau}^2 \Delta^2\right) 
y_{\nu_e}^2\left(1+\frac{2}{3}\Delta^2\right)\left(\frac{\xi}{15}\right)^2
> \la_{3}^2M^4
\label{qcond}
\een
Given \eqn{qcond} and $s^2 > M^2/\lambda_1$, 
we see that our tree potential has no tachyonic instabilities. 

We now proceed to consider the effect of the one-loop corrections 
to the potential. 
Solving the quartic \eqn{quartic}\ exactly 
yields rather unwieldy expressions for the eigenvalues. 
However if we expand the solutions as a series in $a_0$ 
we obtain manageable forms for them as follows: 
\bea
x_{1,2} &=& f_1 \pm \sqrt{d_1}a_0 + e_1 a_0^2 + \cdots\nn 
x_{3,4} &=& f_2 \pm \sqrt{d_2}a_0 - e_1 a_0^2 + \cdots
\label{qusols}
\eea
where 
\bea
f_1&=&\frac{1}{2}\left(a_3-\sqrt{a_3^2-4 a_2 a_1}\right),\nn
f_2&=& \frac{1}{2} \left(a_3+\sqrt{a_3^2-4 a_2 a_1}\right),\nn
d_1&=&\frac{(a_3-a_2-a_1) f_1}{a_3^2-4 a_2 a_1},\nn
d_2&=& \frac{(a_3-a_2-a_1) f_2}{a_3^2-4 a_2 a_1},\nn
e_1&=&\frac{-a_2 (a_3-4 a_1)+a_3 a_1}{2 (a_3^2-4 a_2 a_1)^{\frac{3}{2}}}.
\eea
For each set of four bosonic eigenvalues of the form above, we have
eigenvalues of the corresponding fermion mass matrix of the form
$f_{1,2}$. This is simply because in the absence of $a_0$ (that is to
say, of $M$) the configuration would be supersymmetric.

The contribution to the one-loop potential from 
the matrix \eqn{bigmat}\ becomes (to $O(a_0^2)$), and retaining 
the leading contribution only in each logarithm):

\ben
{16\pi^2}\Delta V_2 = a_0^2\left[ (d_1+ 2f_1
e_1) \ln (f_1/\mu^2) + (d_2 -2f_2 e_1)\ln (f_2/\mu^2) \right].
\een
If we further assume that $a_1 \ll a_{2,3}$ we can simplify $\Delta V_2$ 
by expanding to leading order in $a_1$, when we obtain

\bea
16\pi^2\Delta V_2 &=& \lambda_3^2 M^4 
\left[
1+ M_{F_2}^2
\frac{M_{\nu_1}^2(M_{\nu_1}^2 - M_S^2)}{(M_{\nu_1}^2 + M_S^2)^3}
\right] \ln((M_S^2 + M_{\nu_1}^2)/\mu^2)\nn
&+& \lambda_3^2 M^4 M_{F_2}^2\frac{(M_S^2 - M_{\nu_1}^2) M_{\nu_1}^2}{(M_{\nu_1}^2+M_S^2)^3}
\ln(M_{F_2}^2/\mu^2),
\label{eq:V2}\eea
where we have now written $a_1 \equiv M_{F_2}^2$.
The contribution from the other $(H_1,H_2,Q,E_{1,2},L_3,N_{1,2})$ matrix 
similar to \eqn{bigmat}\ is given by:
\bea
16\pi^2\Delta V_3 &=& \lambda_3^2 M^4 
\left[
1+ M_{F_{1}}^2
\frac{M_{\nu_2}^2(M_{\nu_2}^2 - M_S^2)}{(M_{\nu_2}^2 + M_S^2)^3}
\right] \ln((M_S^2 + M_{\nu_2}^2)/\mu^2)\nn
&+& \lambda_3^2 M^4 M_{F_1}^2\frac{(M_S^2 - M_{\nu_2}^2) M_{\nu_2}^2}{(M_{\nu_2}^2+M_S^2)^3}
\ln(M_{F_1}^2/\mu^2),
\label{eq:V3}\eea
where
\bea
M_{F_1}^2 &=& M_{E_1}^2 +M_D^2 + M_{E_3}^2,\nn
M_{\nu_2} &=& y_{\nu_{\mu}}v_L,\nn
M_{e_1} &=& y_e v_L.
\eea
So our analytic approximation to the scalar potential is finally 
\ben
V = M^4 + \Delta V_1 + \Delta V_2 + \Delta V_3
\label{Vtot}
\een
where $\Delta V_1$ was given in \eqn{Vphiphib}.

\subsection{The $\xi = 0$ case}

In this special case the potential is $D$-flat without invoking the \mssm\ vevs introduced above. 
The one loop potential is dominated by the $\phi,\phib$ system described in section~(\ref{phivevs}), 
and similar contributions from $h_{1,2}$ as is easily seen from \eqn{eq:treepot}. 
For both $\lambda_1 s^2 \gg M^2$ and $\lambda_3 s^2 \gg M^2$ the one-loop corrected potential becomes 
\ben
V = M^4 + 
\frac{1}{16\pi^2}M^4\left[\lambda_1^2 \ln \left(\frac{\la_1^2s^2}{\mu^2}\right)
+ 2\lambda_3^2 \ln \left(\frac{\la_3^2s^2}{\mu^2}\right)\right]. 
\label{Vphih1h2}
\een
This result is easily obtained from \eqn{Vphiphib} and by setting $\xi = 0$ in \eqn{eq:V2} and \eqn{eq:V3}.
(The inequality introduced in \eqn{cond} and \eqn{qcond} is not applicable for $\xi = 0$, 
because this corresponds to $a_2 = a_1 = 0$.) In this case we would require $\lambda_3 > \lambda_1$, 
since otherwise we would find that $s_c^2 = M^2/\lambda_3$, 
rather than $M^2/\lambda_1$, and it would be the Higgses 
that developed vevs rather than $\phi, \phib$. 
Note however that we require $\lambda_3 > \lambda_1$ only if $\lambda_3 \neq 0$; $\lambda_3 = 0$ {\it is\/} 
allowed, since then the Higgs directions which are unstable for $\lambda_3^2 s^2 < \lambda_3 M^2$ become flat.

\section{Inflation}\label{Inflationsection}

In the limit $\la_3 \ll \la_1$ (with $\xi \neq 0$), 
$\Delta V_2$ and $ \Delta V_3$ are
negligible and the effective potential for the $s$ field reduces to that of
standard $F$-term inflation~\cite{Copeland:1994vg}-\cite{Lyth:1998xn}. 
In this section we outline the basic features of this limit as a
reference point, showing how one can estimate constraints from the CMB
data. When we do our more detailed parameter search it will turn out
that we are forced to this limit by other constraints.

The aim is to compute the principal inflationary observables, the scalar and tensor
power spectra $\powspec_s$ and $\powspec_t$, the scalar spectral index $n_s$. 
The importance of the 
tensor power spectrum is often parametrised by $r = 4\powspec_t/\powspec_s$.
In slow-roll single-field inflation these are given by the standard formulae
(see e.g. \cite{Lyth:2009zz})
\bea
\powspec_s(k) & \simeq & \left. \frac{1}{24\pi^2}\frac{V}{m_p^4}\frac{1}{\ep}\right|_{N_k}, \quad
\powspec_t(k) \simeq \left. \frac{1}{6\pi^2}\frac{V}{m_p^4}\right|_{N_k}, \\
n_s & \simeq & \left.(1 - 6\ep + 2\et)\right|_{N_k}, \quad
r = \left.16\ep\right|_{N_k},
\eea
where
\ben
\ep = \frac{m_p^2}{2}\left( \frac{V'}{V}\right)^2, \quad \et = {m_p^2}\left( \frac{V''}{V}\right),
\een
and $N_k$ is the $e$-fold at which the co-moving scale $k$ ``crosses the horizon", i.e.\ $aH=k$. In order to fix $N_{k}$ we need a complete history of the universe, and in particular the temperature to which it reheats after inflation $T_\mathrm{rh}$. When fitting to data 
we are generally interested in the scale $k_0 = 0.002\;\mathrm{ Mpc}^{-1}$, in which case 
$N_{k_0} \simeq 55 + \ln(T_\mathrm{rh}/10^{15}\;\mathrm{GeV})$.

Inflation finishes when $\phi$ and $\bar\phi$ become unstable, 
at a critical value of $s$ given by \eqn{scrit}. From \eqn{phimasses} we see that 
this value for $s$ corresponds to the appearance of a zero eigenvalue 
in the $\phi,\phib$ mass system. 

Let us first consider the case where we approximate $\Delta V_1$ by \eqn{Vphiphib}. This is appropriate 
if $\xi \neq 0$, and $\lambda_3 \ll \lambda_1$. 
If we choose the renormalisation scale $\mu^2 = \la_1^2 s_c^2/2$, we
obtain 
\ben V(s) \simeq M^4 \left[1+ \frac{\lambda_1^2}{16\pi^2}\ln
\frac{2s^2}{s_c^2}\right] \simeq M^4\left(\frac{s_R^2}{s_c^2}
\right)^\alpha, 
\label{eq:vone1}
\een 
where $s_R = \sqrt{2}s$ is a canonically normalised real scalar field, and 
\ben \alpha = \frac{\lambda_1^2}{16\pi^2}.
\een
One can express the solution to the slow-roll equations\footnote{$3H\dot s_R = - V_{,s_R}$, $H^2 = V(s_R)/3\mpl^2$, 
where $H = \dot a/a$ is the 
Hubble parameter and $a$ the cosmological scalar factor.} as
\ben
N(s_R) = \frac{1}{m_p^2}\int_{s_c}^s ds' \frac{V}{V_{,s'}},
\een
where $N = \ln (a_{\rm end}/a(t)) $ is the number of e-foldings before the end of inflation. Hence
\ben
s_R^2 = s_c^2 + 4\al N m_p^2,
\label{eq:sSol}
\een
with 
\ben
\ep = 2\al^2\frac{m_p^2}{s_R^2}, \quad \et = -2\al(1-2\al)\frac{m_p^2}{s_R^2}.
\een
The assumption that $s_R \gg s_c$ is valid provided 
\ben\label{eq:sggsc}
\la_1\al \gg \frac{1}{4N}\frac{M^2}{m_p^2}.
\een
We also want to work in an effective theory well below the Planck scale, 
ensuring $s_R^2 \ll m_p^2$, which is true provided 
\ben\label{eq:sllmp}
\al \ll \frac{1}{4N}.
\een
This is easily satisfied if $\la_1 \ll O(1)$. 

The solution (\ref{eq:sSol}) gives the potential and the 
slow-roll parameters in terms of $N$: 
\bea
\frac{V}{m_p^4} &\simeq& \left(\frac{M}{m_p}\right)^{4-2\al} 
\left(4\al \la_1N_k \right)^\alpha,\\ 
\ep&\simeq& \frac{\al}{2N_k}, \;\;\et \simeq -\frac{1}{2N_k}.
\eea
Hence 
\bea 
\powspec_s(k) & \simeq & \frac{1}{24\pi^2} \frac{2N_k}{\al}
\left(\frac{M}{m_p}\right)^{4-2\al} \left(4\al \la_1N_k \right)^\alpha , \\ 
\powspec_t(k) &\simeq & 
\frac{1}{6\pi^2}\left(\frac{M}{m_p}\right)^{4-2\al} \left(4\al \la_1N_k 
\right)^\alpha, \\ 
n_s & \simeq & \left(1 -
\frac{1}{N_k}\right), 
\eea 
and we find 
\ben
\powspec_s(k) \simeq 7.3
\times 10^{-9} 
\left(\frac{M^2}{10^{-5}\la_1m_p^2}\right)^{2},\quad n_s \simeq 0.982,
\label{eq:powerspec}
\een
where we have taken $N_k = 55$, and neglected a factor raised to the power $\al$, as $\al \sim 10^{-2}$.

The data we we use consists of the WMAP7 best-fit values for $\powspec_s(k_0)$ and $n_s$ in the standard $\La$CDM model are \cite{Komatsu:2010fb}
\ben\label{eq:WMAP7ps}
\powspec_s(k_0) = (2.43\pm0.11) \times 10^{-9}, \quad n_s = 0.963\pm0.012 (68\% \textrm{CL})
\een
with an upper limit on $r$ of \cite{Larson:2010gs}
\ben\label{eq:WMAP7r}
r < 0.36 (95\% \textrm{CL}).
\een

The string contribution is small, so we can equate the $F$-term 
prediction for the scalar power spectrum to 
the WMAP measured value to find 
\ben\label{eq:InfNor}
\frac{M^2}{\la_1m_p^2} \simeq 6 \times 10^{-6}.
\een
Assuming 55 e-foldings of inflation, the allowed range of $\la_1$ is 
therefore approximately
\ben
2.0\times 10^{-3} \ll \la_1 \ll 1,
\label{eq:lam1bound}
\een
where the lower bound comes from the requirement that $s_R\gg s_c$ (\eqn{eq:sggsc}), 
and the upper bound from $s_R^2 \ll \mpl^2$ (\eqn{eq:sllmp}).

Note that the tilt is about $2\sigma$ away from the
best-fit value for single-field inflation. A small string contribution
to the CMB power spectrum at the level of 5-10\% restores or even
slightly improves the CMB fit
\cite{Battye:2006pk,Bevis:2007gh,Battye:2010hg} although at the cost of
a higher baryon fraction and a less steep dark matter power spectrum,
putting the model into tension with other data \cite{Battye:2010hg}.

\subsection{The $\xi = 0$ case}

We can perform a similar analysis to that presented above for the case
$\xi = 0$, $\lambda_3 > \lambda_1$. By exploiting the freedom to add
finite local counterterms to the one loop potential, we may write  (from
\eqn{Vphih1h2}):

\ben V(s) \simeq M^4 \left[1+ \frac{\lambda_1^2 + 2\lambda_3^2}{16\pi^2}\ln
\frac{2s^2}{s_c^2}\right] \simeq M^4\left(\frac{s_R^2}{s_c^2}
\right)^\alpha, \een 
where we still have $s_c^2 = M^2/\lambda_1$, but now 
\ben \alpha = \frac{\lambda_1^2+ 2\lambda_3^2}{16\pi^2}.
\label{eq:newalpha}
\een
The analysis of \eqn{eq:vone1}-\eqn{eq:lam1bound} goes through
essentially unchanged, except that  in
\eqn{eq:powerspec}-\eqn{eq:lam1bound}, $\lambda_1$ is replaced  by
$\sqrt{\lambda_1^2 + 2\lambda_3^2}$.

While we require $\lambda_3 > \lambda_1$, (unless $\lambda_3 = 0$, which
{\it is\/} allowed,  as explained above) we cannot have $\lambda_3 \gg
\lambda_1$, since  were this the case the amplitude of the inflation
perturbations would be proportional to $(M^2/(\lambda_3 m_P^2))^2$ and
dominated by the string perturbations, which are proportional to 
$(M^2/(\lambda_1 m_P^2))^2$.  One sees this easily from
\eqn{eq:powerspec} with $\alpha$ defined by \eqn{eq:newalpha}, and the
string tension by \eqn{eq:StrTen}.

\section{Cosmic Microwave Background string constraints}

The symmetry-breaking of the U(1)$'$ symmetry at the end of inflation 
produces cosmic strings \cite{Hindmarsh:1994re,VilShe94}, which are a
source of gravitational perturbations and contribute to the cosmic
microwave background fluctuations. The exact constraint depends on
details of the modelling of the strings, but simulations of the Abelian
Higgs model compared to WMAP data \cite{Bevis:2007gh}
give
\ben
G\mu \lesssim 7 \times 10^{-7}.
\label{e:Bev+StrCon}
\een
where $\mu$ is the string tension (not to be confused with the
renormalisation  scale $\mu$ appearing in \eqn{eq:AD3}\ or \eqn{Vone},
for example). The Unconnected Segment Model of string perturbations
\cite{Pogosian:1999np} gives a similar upper bound \cite{Battye:2010xz}.
The simple $F$-term hybrid inflation model is more tightly constrained
\cite{Battye:2010hg}, as the string tension is related to the inflation
scale. However, the \amsb{} model has more freedom, and we shall use the
more general string bound  \eqn{e:Bev+StrCon}). For this we will need to
calculate the string tension.

For our model, the string tension is a function of the parameters
$\la_1$, $q_\phi g'$, $M^2$ and $\xi/q_\phi$.  There are two limits
where we can write analytic expressions (see Appendix):

(a) $\xi \gg q_\phi M^2/\la_1$, for which the string tension is 
\ben
\mu_a \simeq 2\pi \xi/q_\phi.
\een

(b) $\xi \ll q_\phi M^2/\la_1$, for which the string tension is 
\ben
\mu_b \simeq 2\pi \tenfun(\la_1^2/2q_\phi^2 g'^2)\frac{2M^2}{\la_1}.
\label{e:StrTenB}
\een
where $\tenfun$ is a slowly varying function of its argument, 
satisfying $\tenfun(1) = 1$. Recall that $q_\phi = 10$ in our model, so its presence is significant.

Case (a) is already ruled out. The string constraint \eqn{e:Bev+StrCon}\
can be rewritten as 
\ben
\frac{\xi}{q_\phi m_p^2} \lesssim 3\times 10^{-6},
\een
which together with \eqn{eq:InfNor}\ is inconsistent with the assumption
$\xi \gg q_\phi M^2/\la_1$. 

Let us turn to case (b). Given that 
$G\mu \lesssim 7 \times 10^{-7}
$
we can substitute the string tension \eqn{e:StrTenB}\ and 
use the inflationary normalisation \eqn{eq:InfNor}\ to 
derive an approximate upper bound on $\tenfun$, 
\ben
\tenfun \lesssim 
7 \times 10^{-7}\left(\frac{2\la_1m_p^2}{M^2}\right) 
\simeq 0.2.
\een
Hence the value of $\la_1^2/2q_\phi^2 g'^2$ has to be small in order for
strings not to exceed the CMB bound. Using the approximation 
\cite{Hill:1987qx} $\tenfun(\beta) \simeq 2.4/\ln(2/\beta)$ for $\beta
\lesssim 10^{-2}$, we find  \ben \frac{\la_1}{\sqrt{2}q_\phi g'}
\lesssim 3\times 10^{-3}. \een

A Monte-Carlo fit in the simple $F$-term inflation model, using the
numerically determined string tension, and taking into account the
degeneracies between $G\mu$ and the other cosmological parameters, has
been performed by Battye, Garbrecht \& Moss \cite{Battye:2010hg}. 
One should note that their inflation superpotential is $W_{\rm BGM}
= \ka S(\Phi\bar\Phi - M_{\rm BGM}^2)$, so that  $\kappa=\la_1$ and
$M_{\rm BGM} = M/\sqrt{\la_1}$, and that they take $g'=0.7$ and $q_\phi
= 1$. They take the string tension to be~\cite{Jeannerot:2006jj}
\eqn{eq:AHstrTen}, with $v^2 = M_{\rm BGM}^2$.  The best fit models have
$M_{\rm BGM}\sim 5 \times 10^{15}$ GeV and $\ka \sim 10^{-3}$ --
$10^{-2}$.

Comparison is not straightforward, as our model's string tension receives
contributions from the $D$-term \eqn{eq:StrTen}.\footnote{Note that
the two formulae disagree by a factor 2 even in the limit $\xi \to 0$.
This is the result of an incorrect application 
(in \reference{Battye:2010hg}) of the standard formula for
Abelian Higgs string tension (\ref{eq:AHstrTen}), by neglecting the
fact that $F$-term strings have two scalar fields.} This difference in
the string tension formulae, the accuracy of the fit, and the slowly
varying nature of the function $B$ mean that our estimate on
the coupling ($\la_1 \lesssim 0.4$) is broadly compatible with the upper
bound on their $\ka$.

\section{Numerical scan of parameter space}

We have done some numerical testing of the parameter space of the model,
looking for combinations which are consistent with our assumptions and
the data. We have designated the \sm{} parameters to their measured
value, taken $\tan(\beta)=60$, and, as mentioned, taken the Yukawa
couplings to be real and diagonal. In the subspace we have investigated
this leaves the following 6 variables:
$M,\la_1,\la_3,y_{\nu_e},y_{\nu_\mu},\sqrt{\xi}$. In this analysis we
used the approximations for the appropriate mass eigenvalues given by
\eqn{qusols}. The following conditions must be satisfied for a
successful model.

\begin{enumerate}

\item All vevs, masses and mass scales should be less than the Planck
scale, otherwise our neglect of gravitational corrections becomes
inconsistent. We check the value of the inflation field $s$ and all
masses between $s=s_{55}$, where $s_{55}$ is the $s$ field value 55
e-foldings before the end of inflation, and $s=s_{c}$.

\item The string tension should satisfy the CMB upper bound
\eqn{e:Bev+StrCon}, which we use in conjunction with the formulae of
the appendix. This depends on $g'$ through \eqn{e:BetaEff}; we have
taken a weak limit, namely the one that arises from $g'=1$. We have not
treated $g'$ as a variable since there is no dependence on $g'$ other 
than here in the string tension.

\item There should be no tachyons during inflation, other than those
which drive the vevs of $\phi$ and $\bar\phi$ at the end of inflation. 
This means obeying \eqn{qcond}. 

\item The amplitude of scalar perturbations should be consistent with 
observations, \eqn{eq:WMAP7ps}. We ignore the small string contribution
to the power spectrum for the purpose of our approximate survey.

\item We require that the scalar spectral tilt $n_s$ be within 3$\si$ of
its measure mean value, \eqn{eq:WMAP7ps}.

\item We require that all couplings be perturbative, i.e.\ less than 1
at the inflation scale $M$. Applying this condition (approximately) to
the elements of the matrix $\la_2$ puts an upper bound on the neutrino
Dirac Yukawa couplings, via the seesaw formula
\ben\label{eq:Neutmass}
m_{\nu_i}=\frac{m_D^2}{m_N}=\frac{y_{\nu_i}^2
v^2 \sin^2\beta}{\la_{2i,\rm eff}v_{\phi}}.
\een
Here $v = 174\GeV$ and $m_{\nu}$, $m_D$, $m_N$ are the physical, Dirac
and Majorana masses of the neutrino, and $\la_{2i,\rm eff}$ is the
effective parameter from $\la_2$ relevant for the $i$'th neutrino state.
The vev $v_{\phi}$ is determined by Eqs.~(\ref{vmina})-(\ref{vminc}).
The CMB upper limit on the sum of the neutrino masses of approximately
1.5 eV \cite{Komatsu:2010fb} can then be translated into an upper bound
$y^2_{\nu_i} < m_{\nu_i}v_\phi/v^2\sin^2\beta$.  In practice this does
not set an extra bound on $y_{\nu_i}$ above the perturbativity bound, as
$m_{\nu_i}v_\phi/v^2 \lesssim 50$.

\end{enumerate}

Some successful parameter combinations are placed in the left half of
Table~\ref{NumResTab}. In the right half we put the resulting slow-roll
parameters, scalar amplitude, tilt, string tension as a fraction of the
$\mu_{max}$ (the maximal tension Eq.\ \ref{e:Bev+StrCon} evaluated with
$g'=1$), and the value of the inflaton in Planck units at $N = 55$
e-folds of inflation.

The first line is the case when $\la_3=0$ 
and $\xi = 0$, which is the limit in which
the radiative corrections from the \mssm{} field vanish, $\De V_2 = \De
V_3 = 0$, and the system reverts to the simple F-term model of section
\ref{sec:quatics}. 
This case seems to have a high $n_s$ compared to the single-field inflation
mean value of 0.963 \cite{Komatsu:2010fb}, but as pointed out in
Section \ref{Inflationsection}, a high $n_s$ is a feature in models 
with cosmic strings contributing 5-10\% to the power spectrum at $\ell
= 10$ \cite{Battye:2006pk,Bevis:2007gh}. This scenario ($\la_3 = 0$) 
would mean we would require an alternative source for the Higgs $\mu$-term. 
The second row is another case with vanishing 
FI term $\xi = 0$, but $\la_3 \ne 0$.

For the other cases, we have
demanded $\la_1\la_3\geq 10^{-5},\sqrt{\xi}/M\geq 0.2$ to reject values
of the effective $\mu$ that are unnecessarily low and so that $\xi,\la_3$
actually play a role. We see that $n_s$ is almost independent of
$\xi,\la_3$ with the assumptions made.

These parameter values should only be taken as indicative of regions of
parameter space where a proper Monte Carlo fit with more accurate
formulae should be undertaken. This we leave for a future work.

The third row has the highest $\la_1\la_3$ and $\la_3$, the fourth has
the highest $n_s$, the fifth has the lowest $n_s$ and lowest
$M^2/\la_1$, the sixth has the highest $M^2/\la_1$ and the seventh has the
highest $\sqrt{\xi}$. 

In a supergravity extension of the model, with K\"ahler potential 
\ben
K = |s|^2 + c|s|^4/\mpl^2,
\een
the tree-level potential would be modified to 
\ben
V_{\rm sugra} = 
M^4\left( 1 - 4c \frac{|s|^2}{\mpl^2}
+  \left(\frac{1}{2}-7 c + 16 c^2\right)\frac{|s|^4}{\mpl^4}\right),
\een
which potentially gives rise to the well-known supergravity
$\eta$-problem \cite{Copeland:1994vg}. One can hope that some symmetry
sets $c = 0$, but a quartic term in the scalar potential is unavoidable, 
and a minimal requirement for success of the model is that 
$s_{55}/ \mpl \ll 1$.  
One can easily check which cases
are afflicted by the $\eta$-problem 
by evaluating 
\ben
\De \et_{\rm sugra} = \mpl^2 \frac{V''_{\rm sugra}}{V_{\rm sugra}},
\een
which is an upper bound on the change in $\eta$ due to the supergravity
corrections.  Recall that  $s_R = (\Re s)/\sqrt{2}$, and without loss of
generality we may suppose that $s$ is real. In the $c=0$ case we find
that
\ben
\De \et_{\rm sugra} \simeq 
\frac32 \frac{s_R^2}{\mpl^2}\left(1+\frac18\frac{s_R^4}{\mpl^4}\right)^{-1}.
\een
Using the values of $s_{55}/\mpl$ listed in the last column of the 
table, we can verify that only the first is afflicted, with 
$\De \et_{\rm sugra} = 0.122$. 

A possible (albeit tuned) solution is to posit a small positive value of
$c$, which reduces $\eta$ by an amount $\Delta \eta \sim - 4c$. The
supergravity corrections can also be reduced by reducing $s_{55}$, and 
from \eqn{eq:sSol} one sees this can be effected with smaller values of
$\la_1$.   If $M^2$ is also reduced, so as to keep $M^2/\la_1$ constant,
then the string tension $\mu$ decreases approximately logarithmically. 

The problem, however, lies in the requirement that, although 
$\la_3 \geq \la_1$, we cannot have $\la_3 \gg \la_1$,  because then, as
mentioned  in section~6.1, the inflation CMB perturbations would be 
suppressed relative to the string ones. Inevitably, therefore, the   
Higgs $\mu$-term $\mu_h = \la_3 v_s$ would be reduced.  Here we thus
encounter an interesting tension between the requirements of the theory
in two quite different epochs, corresponding to inflation and 
electro-weak symmetry breaking.

\begin{table}
\begin{center}
\begin{tabular}{|c c c c c | c c c c c  c |} \hline
 $\frac{M}{10^{14}\GeV}$ & $\frac{\la_1}{10^{-3}}$
& $\frac{M^2}{10^{-6}\la_1m_{pl}^2}$ & $\frac{\sqrt{\xi}}{M}$&
 $\frac{\la_3}{10^{-3}}$ & $\frac{\ep}{10^{-9}}$
& $\frac{\eta}{10^{-3}}$&
 $\frac{{\cal P}_s}{10^{-9}}$& $n_s$& $\frac{\mu}{\mu_{max}}$ 
& $\frac{s_{55}}{10^{-3}\mpl}$ \\ \hline
 $23.1$ & $161$& $5.62$ & $0 $& $0 $& $5720$& $-8.70$&
 $2.43$& $0.983$& 
 $1.000$ & 190 \\ \hline
$5.28$ & $4.87$& $9.71$ & $0 $& $4.87 $& $3550$& $-7.89$&
 $2.43$& $0.984$& 
 $1.000$ & 10.7 \\ \hline
$7.98$ & $18.3$& $5.89$ & $0.284 $& $4.33 $&
 $20.3$& $-8.61$&
 $2.41$& $0.983$& 
 $0.993$ & 23.3 \\ \hline
$4.12$ & $4.31$& $6.67$ & $0.287 $& $2.61 $&
 $1.42$& $-5.66$&
 $2.45$& $0.989$& 
 $0.981$  & 6.54 \\ \hline
$13.0$ & $51.0$& $5.61$ & $0.200 $& $0.196 $&
 $1440$& $-8.75$&
 $2.41$& $0.982$& 
 $0.987 $ & 61.5 \\ \hline
$4.12$ & $4.03$& $7.13$ & $0.200 $& $2.61 $&
 $1.42$& $-6.41$&
 $2.44$& $0.987$& 
 $0.983$ & 6.65 \\ \hline
$4.12$ & $5.10$& $5.64$ & $0.590 $& $2.38 $&
 $1.42$& $-6.77$&
 $2.44$& $0.986$& 
 $0.962 $ & 6.78 \\ \hline
 
\end{tabular}
\caption{\label{NumResTab}
Mass scale and coupling constant values (left half), with inflationary
and  cosmic string parameters (right half) for cases satisfying the
numbered constraints in the text.}
\end{center}
\end{table}

Note that values of $\lambda_{1,3}$ as small as those obtained above
have consequences for the Higgs $\mu$ term. This was given by $\mu =
\la_3 v_s$; if we assume that $\rho$ is, unlike the other soft terms,
unsuppressed by loop factors, we find from \eqn{vminc} that $v_s \sim
\rho \lambda_1 m_{\frac{3}{2}}$ suggesting that $\mu \sim O(\GeV)$
rather than $O(100\GeV)$. This will impact the electroweak  vacuum
minimisation and the associated sparticle spectrum, which we will
explore elsewhere\footnote{A constraint on $\mu_h$ might, for example,
lead  to a prediction for $\tan\beta$ from the minimisation.}.

\section{Other Cosmological constraints}

While the main topic of this paper is the possibility for  inflation in
\amsb, there are other cosmological phenomena to consider. They come
both as bounds and as possibilities to  explain observed phenomena, e.g.
the presence of baryonic and dark matter. \subsection{Gravitational wave
and cosmic ray constraints}

Cosmic strings have a property known as \textit{scaling}
\cite{Hindmarsh:1994re,VilShe94}, which means that they maintain a
constant density parameter $\Om_s= \rho_s/\rho_c$, where $\rho_s$ is the
string energy density and $\rho_c = 3\mpl^2H^2$ is the critical density.
Numerical simulations~\cite{StringSims}-\cite{Hindmarsh:2008dw} indicate
that there are $O(1)$  or a few Hubble lengths of string per Hubble
volume, so that $\rho_s \sim \mu/t^2$ and $\Om_s \sim G\mu$.

The string energy density is therefore decaying at a rate $\dot\rho_s
\sim \mu/t^3$. There are two scenarios for the products of this decay:
the primary channel is either via closed loops of string into
gravitational radiation or into high energy particles of the fields from
which the string is made. The first scenario is constrained by bounds on
the stochastic background of gravitational radiation (see
e.g.~\cite{Siemens:2006yp}), and the second by the flux of cosmic rays
\cite{Vincent:1997cx}. The first depends on the typical loop size
relative to the Hubble length, $\al$, and the second on the complex
decay processes of the massive scalars, gauge bosons, and fermions of
the string fields, here the $\phi$, $\bar\phi$, $U(1)'$ gauge field, and
the neutrino zero modes. It is also possible that the Higgs field $h_2$
has a vev in the string core (see Appendix).

In the first scenario, with the assumption about the average loop size
$\al \ll G\mu$, the upper bound on the string tension is
\cite{Battye:2010xz}
\ben
\label{e:GmuPulsarBound}
G\mu \lap
7 \times 10^{-7 }.
\een 
In the second scenario a detailed modelling of the decay cascades is
required as the bounds on the mass scale are sensitive to the primary
Standard Model decay products of the ``X" particles into which the
string decays \cite{Bhattacharjee:1998qc}.  Cosmic strings constitute a
$p=1$ TD (topological defect or top-down) \cite{Bhattacharjee:1998qc}
model,  for which  there is a upper bound on the energy injection rate
from the low energy diffuse $\ga$-ray background\cite{Bhattacharjee:1998qc} 
\ben
Q_0 \lesssim 4.4\times 10^{-23}h \; \textrm{eV cm$^{-3}$ s$^{-1}$}.
\een
Assuming the massive particles decay into Standard Model particles with
a non-zero branching fraction $f$,  one can derive a
bound~\cite{CRconstraints}-\cite{Wichoski:1998kh}
\ben
G\mu \lesssim 10^{-9} x_*^2f^{-1},
\een
where $x_*^2 = \mu/\rho_st^2$ parametrises the Hubble lengths of string
per Hubble volume,  with $x_* \sim 0.3$
\cite{StringSims}-\cite{Hindmarsh:2008dw}. The string gauge field 
couples to all Standard Model particles, so the cosmic ray constraints
are potentially strong in this scenario. In view  of the uncertainty
about which is correct, we have adopted the weaker constraint
(\ref{e:GmuPulsarBound}), which is no stronger than the CMB bound
(\ref{e:Bev+StrCon}).

\subsection{Dark Matter}

Low energy supersymmetry (with the imposition of $R$-parity) has the
attractive feature that the \lsp\ is stable, and is thus a dark matter
candidate. For a recent general review of the situation, see
\cite{Ellis:2010kf}. For an agnostic, the parameter space of low energy
\sy\ allows many candidates for the \lsp, including the gravitino; in
conventional \amsb{} the gravitino is certainly too heavy. Since a
charged \lsp\ would surely have been detected in terrestrial studies, a
framework which automatically excludes them is to be welcomed. This is
true of the version of \amsb{} we discuss here, except very close to the
boundaries of $(q_L,q_E)$ parameter space, where the \lsp\ can in fact
be a charged lepton. Generally, however, the \lsp\ is a neutralino with
a  dominant neutral wino component~\cite{Hodgson:2005en}.  This has been
argued to disfavour it as a dark matter candidate in \amsb, because of
the (comparatively) large annihilation cross-section of such a
neutralino~\cite{Giudice:1998xp}. However, \amsb{} models in general,
and our model in particular,  could produce the required neutralino
abundance from the decays of thermally produced gravitinos, provided
that the reheat temperature is high
enough~\cite{Gherghetta:1999sw}-\cite{Baer:2010kd}.  For $m_{\frac32}
\simeq 40$ TeV this is around $2\times10^{10}$ GeV.  (For a Bayesian
analysis of how the minimal \amsb{} scenario is constrained by other 
observables, see~\cite{AbdusSalam:2009tr}). 

Our model also has cosmic strings, and neutralinos will generically be
produced by decays of particles  radiated from
them~\cite{Jeannerot:1999yn}.  We note that this and other works (see
for example~\cite{DMfromStrings,Bi:2009am})  studying dark matter
production from strings make conservative assumptions about the amount
of particle production by assuming that gravitational radiation
dominates, so a re-calculation of the dark matter density as a function
of the string tension $\mu$ and the branching fraction of string decays
into neutralinos would be extremely useful.

Finally, one might also entertain the possibility that the \amsb{}
pattern of supersymmetry breaking is associated with a mass scale other
than the gravitino mass; we would then be free to consider a gravitino
light enough to be the \lsp, with the wino-dominated neutralino now the
\nlsp\ and metastable. That would however, not be consistent with the
leptogenesis scenario described in the next section. 

\subsection{Baryogenesis}

Creation of the observed baryon asymmetry requires baryon number
violation,  departure from thermal equilibrium  and C and
\cp-violation~\cite{Sakharov:1967dj}. Our model has no conserved lepton
numbers, and so it is natural  to explore creation of the observed
baryon number via leptogenesis.

Now in \amsb, non-\ckm\ \cp-violating phases do not exist in the
soft-breaking sector, apart from a possible phase associated with the
$\mu$ and $B$ terms, for which we do not have a complete theory; they
are simply constrained so as to produce the \sm{} vacuum.  To put it
another way, $\kappa$ in \eqn{eq:AD4}\ could be  complex. An interesting
potential  source (of \cp\ violation) is, however, the Yukawa sector for
right-handed neutrinos, which (with the standard see-saw mechanism for
generating neutrino masses) is  relevant for
leptogenesis~\cite{Fukugita:1986hr}. Successful supersymmetric
leptogenesis requires that the lightest right-handed neutrino (and the
post-inflation reheating temperature) be greater than $10^9\GeV$; note
that because of the large gravitino mass (around $40\TeV$)  associated
with \amsb{} there is no danger that the decay of gravitinos produced in
this reheating will pose a problem for 
nucleosynthesis~\cite{Ibe:2004tg}. (For a recent discussion of some
other ways of evading the gravitino bound see
\reference{Giudice:2008gu}).

There is also a source of leptogenesis through out-of-equilibrium decays
of particles radiated by the strings, along the lines of the scenario
investigated in Refs.~\cite{Jeannerot:1996yi}-\cite{Sahu:2005vu}\  for
$B-L$ cosmic strings. It would be interesting to investigate this
further in conjunction with the dark matter and cosmic ray constraints
on strings.

Finally, in the $\xi \neq 0$ case, our \amsb{} model also has the right
conditions for Affleck-Dine baryogenesis \cite{Affleck:1984fy}, in that
inflation naturally generates large vevs for fields with baryon and
lepton number through the minimisation of the $D$-term. A detailed
investigation would involve numerical simulations of the dynamics of the
fields at the end of inflation, which is beyond the scope of this paper.

\section{Conclusions}

We have shown how a theory with low energy \sy{}, constructed so as to
produce a viable sparticle spectrum based on anomaly mediation, also has
significant cosmological consequences. The \amsb{} scenario is an
attractive alternative to (and easily distinguished from) the \cmssm.
(It was believed that \amsb{} was disfavoured in terms of accommodating
the existing discrepancy between theory and experiment for the anomalous
magnetic moment of the muon; but this conclusion has been challenged
recently~\cite{Allanach:2009ne}). We have shown how a $U(1)'$ gauge
symmetry originally introduced to solve the \amsb{} tachyonic slepton
problem leads to interesting cosmological possibilities. 

In the minimal form presented here, the $U(1)'$ gauge symmetry
requires three extra chiral Standard Model singlets, two of which are
charged under the $U(1)'$. From this new structure we obtains a
$\mu$-term, Majorana masses for the right-handed neutrinos, 
and potentially \cp-violating mixings.

The model naturally realises $F$-term hybrid inflation, terminating with
the production of cosmic strings. CMB data put strong constraints on
the extra parameters introduced, principally the $F$-term and $D$-term
mass scales $M$ and $\sqrt{\xi}$, and the inflaton couplings $\la_1$ and
$\la_3$. If $\xi \neq 0$, the $D$-term induces squark and slepton vevs during
inflation, which allows Affleck-Dine baryogenesis to take place, using
\cp-violation in the neutrino sector. If we set $\xi = 0$, as argued for in 
\reference{Komargodski:2009pc}, then there are other sources of baryogenesis
include conventional leptogenesis and non-thermal leptogenesis from
cosmic string decays. Cosmic string and gravitino decays also boost the
dark matter density, which is normally low in the conventional
freeze-out scenario.

We have seen that choosing parameters so as to avoid the
$\eta$-problem  has the surprising consequence that the prediction for
the Higgs $\mu$-term  is reduced. If it proves nevertheless possible to
implement electro-weak breaking  in a satisfactory way this will count
as a success for the model, providing  as it does a potential solution
for the ``little hierarchy'' problem.

In conclusion, our \amsb{} model can satisfy the principal cosmological
constraints, and provide an acceptable particle physics phenomenology,
in the framework of a renormalisable quantum field theory with few extra
parameters above those of the Standard Model. There are the neutrino
coupling matrices $\la_2$ and $Y_N$ (which are common to models
incorporating neutrinos), two coupling constants $\la_1$ and $\la_3$ and
the leptonic  $U(1)'$ charges $(q_L,q_E)$. There are in general two mass
scales $M$ and $\xi$,  and two parameters associated with the
supersymmetry breaking, $m_{\frac32}$ and $\ka$. This economy makes this
a model worthy of more detailed investigation on all fronts.

\section*{\large Acknowledgements}

While part of this work was done, DRTJ was at the  Centro de Ciencias de
Benasque Pedro Pascual, and also at KITP Santa Barbara. This research
was supported in part by the National Science Foundation  under Grant
No. NSF PHY05-51164, and by the Science and Technology Research Council
[grant numbers ST/G000573/1 and ST/G00062X/1]. 
AB was supported by The Danish Council for
Independent Research $|$ Natural Sciences. 

\appendix

\section{String tension in the \amsb{} model}
For the ordinary Abelian Higgs model the string tension is given 
in terms of the quartic coupling $\la$, 
the gauge coupling which we will call $g'$, and the expectation value of 
the complex scalar field $v$. With the ansatz
\ben
\phi(r,\th) = v R(r)e^{i\th},\qquad A_i = \hat\th_i \frac{a(r)}{g'r},
\een
and boundary conditions $a,R\to 0$ as $r\to 0$ and $a,R\to 1$ as $r\to \infty$
we obtain a solution for which the string tension is
\ben
\mu = 2\pi \int_0^\infty dr\, r \left( \half \left(\frac{a'}{g'r}\right)^2 
+ v^2(R')^2 + \frac{(1-a)^2}{r^2} R^2 v^2 + \half \la v^4(R^2 - 1)^2 \right)
\een
Defining a dimensionless radial coordinate $z = rvg'$, we have
\bea
\mu &=& 2\pi v^2 \int_0^\infty dz\, z 
\left( \half \left(\frac{a'}{z}\right)^2 + (R')^2 
+ \frac{(1-a)^2}{r^2} R^2 
+ \half \beta(1 -R^2)^2 \right)
\eea
where $\beta = \la/{g'}^2$. In the special case $\beta=1$, the string
tension is $2\pi v^2$. More generally 
\ben\label{eq:AHstrTen}
\mu = 2\pi \tenfun(\beta) v^2,
\een
where $\tenfun$ is a slowly varying function of its argument, 
satisfying $\tenfun(1) = 1$. For low $\beta$ the function can be
approximated by  \cite{Hill:1987qx}
\ben
B(\beta) \simeq \left\{ 
\ba{cc}
1.04\beta^{0.195}, & 10^{-2} < \beta \ll 1 \cr 
2.4/\ln(2/\beta), & \beta < 10^{-2} 
\ea \right.
\een
For the \amsb{} model, the string tension is a function of the
parameters $\la_1$, $q_\phi g'$, $M^2$ and $\xi/q_{\phi}$.  The ansatz is 
\ben
\phi(r,\th) = v_\phi R(r)e^{i\th},\qquad 
\bar\phi(r,\th) = -v_{\bar\phi} \bar R(r)e^{-i\th}, \qquad A_i = \hat\th_i \frac{a(r)}{q_\phi g'r},
\een
with the vacuum expectation values of the fields $\phi$ and $\bar\phi$ 
given by (see~Eqs.~(\ref{vmina}),(\ref{vminb}))
\bea
v_\phi^2 &=& \frac12 \left[ \sqrt{\left(\frac{\xi}{q_\phi}\right)^2 + \left( \frac{2M^2}{\la_1}\right)^2} + \frac{\xi}{q_\phi}\right], \\
v_{\bar\phi}^2 &=& \frac12 \left[ \sqrt{\left(\frac{\xi}{q_\phi}\right)^2 + \left( \frac{2M^2}{\la_1}\right)^2} - \frac{\xi}{q_\phi}\right].
\eea
The string tension is 
\bea
\mu &=& 2\pi \int_0^\infty dr\, r \left( \half \left(\frac{a'}{q_\phi g'r}\right)^2 
+ \left[v_\phi^2(R')^2 + v_{\bar\phi}^2(\bar R')^2\right] 
+ \frac{(1-a)^2}{r^2} \left[v_\phi^2R^2 + v_{\bar\phi} ^2\bar R^2\right] \right.\nonumber\\
&& \left.+ \half g'^2 (\xi -q_\phi v_\phi^2 R^2+q_\phi v_{\bar\phi} ^2 \bar R^2)^2 + (\la_1 v_\phi v_{\bar\phi}R \bar R - M^2)^2\right). 
\label{e:StrTenFun}
\eea
We can get an upper bound and a reasonable approximation by assuming that $\bar R = R$, in which 
case the string tension can be written 
\bea
\mu &\simeq& 2\pi (v_{\phi}^2 + v_{\bar\phi}^2)\int_0^\infty \hspace{-3pt}dz \,z 
\left[ \half \left(\frac{a'}{z}\right)^2 + (R')^2 + \frac{(1-a)^2}{z^2} R^2 + \frac{\beta_{\rm eff}}{2}(1 -R^2)^2 \right]
\eea
where
\ben
z^2 = q_\phi^2 {g'}^2r^2(v_\phi^2+v_{\bar\phi}^2), \quad \beta_{\rm eff} = \frac{1 + \psi \beta}{1 + \psi}.
\een
Here we have defined
\ben\label{e:BetaEff}
\psi = \frac{2M^2q_\phi}{\la_1 \xi}, \quad \beta = \frac{\la_1^2}{2q_\phi^2 {g'}^2}.
\een
Thus we see that the string tension in the \amsb{} model is approximately
\ben \label{eq:StrTen}
\mu \lesssim 2\pi B(\beta_{\rm eff} ) \sqrt{\left(\frac{\xi}{q_\phi}\right)^2+ \left( \frac{2M^2}{\la_1}\right)^2}.
\een
The approximation becomes an equality in the limits $\psi \to 0, \infty$. In the first case the assumed symmetry between $\phi$ and $\bar\phi$ becomes 
exact as the $D$-term becomes negligible, and in the second case $\bar\phi$ vanishes as the $F$-term becomes negligible.
The expressions for the string tension is these two limits is 
\ben\label{eq:stringsimple}
\mu_a = 2 \pi \frac{\xi}{q_\phi}, \quad \mu_b = 2\pi B(\beta) \frac{2M^2}{\la_1} .
\een
A more accurate solution can be obtained by a numerical minimisation of
the string tension function (\ref{e:StrTenFun}). One should also
allow for the possibility of the \mssm{} scalars with positive $q_\phi$
gaining an expectation value in the core of the string, as this reduces
the $D$-term potential energy density which would otherwise be $g'^2/\xi$
at the core of the string. A prime candidate is the Higgs field $h_2$,
as it already has a vev. The other candidates are $l$ and $e^c$, but
they have lower $q_\phi$ and are therefore less unstable in the string
core.

Finally, we note that the string will have fermionic zero modes from two
sources: from the neutrinos thanks to the
$\half\la_2 N N \Phi$ coupling, and from mixtures of the superpartners of
$s$, $\phi$ and $\bar\phi$, thanks to the
$S\Phi\bar\Phi$ coupling~\cite{Davis:1997bs}.


\begin{thebibliography}{99}
\bibitem{Randall:1998uk}
 L.~Randall, R.~Sundrum,
 Nucl.\ Phys.\ {\bf B557 } (1999) 79-118.
 [hep-th/9810155].
\bibitem{Giudice:1998xp}
 G.~F.~Giudice, M.~A.~Luty, H.~Murayama {\it et al.},
 JHEP {\bf 9812 } (1998) 027.
 [hep-ph/9810442].
\bibitem{Pomarol:1999ie}
 A.~Pomarol, R.~Rattazzi,
 JHEP {\bf 9905 } (1999) 013.
 [hep-ph/9903448].
\bibitem{Jack:2000cd}
 I.~Jack, D.~R.~T.~Jones,
 Phys.\ Lett.\ {\bf B482 } (2000) 167-173.
 [hep-ph/0003081].
\bibitem{Arkani-Hamed:2000xj}
N.~Arkani-Hamed, D.~E.~Kaplan, H.~Murayama and Y.~Nomura,
JHEP {\bf 0102} (2001) 041.
[hep-ph/0012103].
\bibitem{Harnik:2002et}
R.~Harnik, H.~Murayama and A.~Pierce,
JHEP {\bf 0208} (2002) 034.
[hep-ph/0204122].
\bibitem{Murakami:2003pb}
 B.~Murakami, J.~D.~Wells,
 Phys.\ Rev.\ {\bf D68 } (2003) 035006.
 [hep-ph/0302209].
\bibitem{Kitano:2004zd}
R.~Kitano, G.~D.~Kribs and H.~Murayama,
Phys.\ Rev.\ {\bf D70} (2004) 035001.
[hep-ph/0402215].
\bibitem{Ibe:2004gh}
M.~Ibe, R.~Kitano and H.~Murayama,
Phys.\ Rev.\ {\bf D71} (2005) 075003.
[hep-ph/0412200].
\bibitem{Jack:2002pn}
 I.~Jack, D.~R.~T.~Jones, R.~Wild,
 Phys.\ Lett.\ {\bf B535 } (2002) 193-200.
 [hep-ph/0202101].
\bibitem{Hodgson:2005en}
 R.~Hodgson, I.~Jack, D.~R.~T.~Jones {\it et al.},
 Nucl.\ Phys.\ {\bf B728 } (2005) 192-206.
 [hep-ph/0507193].
\bibitem{Jones:2006re}
 D.~R.~T.~Jones, G.~G.~Ross,
 Phys.\ Lett.\ {\bf B642 } (2006) 540-545.
 [hep-ph/0609210].
\bibitem{Witten:1981kv}
 E.~Witten,
 Phys.\ Lett.\ {\bf B105 } (1981) 267.
\bibitem{Battye:2010hg}
 R.~Battye, B.~Garbrecht and A.~Moss,
 Phys.\ Rev.\ {\bf D81} (2010) 123512.
 [arXiv:1001.0769 [astro-ph.CO]].
\bibitem{Garbrecht:2006}
 B.~Garbrecht and A.~Pilaftsis,
 Phys.\ Lett.\ B {\bf 636} (2006) 154-165.
 [arXiv:hep-ph/0601080]. 
%
\bibitem{Garbrecht:2006az}
 B.~Garbrecht, C.~Pallis and A.~Pilaftsis,
 JHEP {\bf 0612} (2006) 038.
 [arXiv:hep-ph/0605264].

\bibitem{Komargodski:2009pc}
 Z.~Komargodski, N.~Seiberg,
 JHEP {\bf 0906 } (2009) 007.
 [arXiv:0904.1159 [hep-th]].

\bibitem{Dienes:2009td}
 K.~R.~Dienes, B.~Thomas,
 Phys.\ Rev.\ {\bf D81 } (2010) 065023.
 [arXiv:0911.0677 [hep-th]].


\bibitem{Komargodski:2010rb}
 Z.~Komargodski, N.~Seiberg,
 JHEP {\bf 1007 } (2010) 017.
 [arXiv:1002.2228 [hep-th]].

\bibitem{Distler:2010zg}
 J.~Distler, E.~Sharpe,
 [arXiv:1008.0419 [hep-th]].


\bibitem{Banks:2010zn}
 T.~Banks, N.~Seiberg,
 [arXiv:1011.5120 [hep-th]].





\bibitem{Allanach:2009ne}
 B.~C.~Allanach, G.~Hiller, D.~R.~T.~Jones and P.~Slavich,
 JHEP {\bf 0904} (2009) 088.
 [arXiv:0902.4880 [hep-ph]].


\bibitem{Weinberg:1981wj}
 S.~Weinberg,
 Phys.\ Rev.\ {\bf D26 } (1982) 287.
\bibitem{Dimopoulos:1981dw}
 S.~Dimopoulos, S.~Raby, F.~Wilczek,
 Phys.\ Lett.\ {\bf B112 } (1982) 133.
\bibitem{Sakai:1981pk}
 N.~Sakai, T.~Yanagida,
 Nucl.\ Phys.\ {\bf B197 } (1982) 533.
\bibitem{Jack:2001ew}
 I.~Jack, D.~R.~T.~Jones and R.~Wild,
 Phys.\ Lett.\ {\bf B509} (2001) 131.
[arXiv:hep-ph/0103255].

\bibitem{Ellwanger:2009dp}
 U.~Ellwanger, C.~Hugonie, A.~M.~Teixeira,
 Phys.\ Rept.\ {\bf 496 } (2010) 1-77.
 [arXiv:0910.1785 [hep-ph]].
\bibitem{Copeland:1994vg}
 E.~J.~Copeland, A.~R.~Liddle, D.~H.~Lyth {\it et al.},
 Phys.\ Rev.\ {\bf D49 } (1994) 6410-6433.
 [astro-ph/9401011].

\bibitem{Dvali:1994ms}
 G.~R.~Dvali, Q.~Shafi, R.~K.~Schaefer,
 Phys.\ Rev.\ Lett.\ {\bf 73 } (1994) 1886-1889.
 [hep-ph/9406319].

\bibitem{Lyth:1998xn}
 D.~H.~Lyth, A.~Riotto,
 Phys.\ Rept.\ {\bf 314 } (1999) 1-146.
 [hep-ph/9807278].

\bibitem{Binetruy:1996xj}
 P.~Binetruy and G.~R.~Dvali,
 Phys.\ Lett.\ {\bf B388} (1996) 241.
 [arXiv:hep-ph/9606342].

\bibitem{Lyth:2009zz}
 D.~H.~Lyth, A.~R.~Liddle,
 Cambridge, UK: Cambridge Univ. Pr. (2009) 497 p

\bibitem{Komatsu:2010fb}
 E.~Komatsu {\it et al.} [WMAP Collaboration],
 [arXiv:1001.4538 [astro-ph.CO]]

\bibitem{Larson:2010gs}
 D.~Larson, J.~Dunkley, G.~Hinshaw {\it et al.},
 [arXiv:1001.4635 [astro-ph.CO]].

\bibitem{Battye:2006pk}
 R.~A.~Battye, B.~Garbrecht and A.~Moss,
 JCAP {\bf 0609} (2006) 007.
 [arXiv:astro-ph/0607339].

\bibitem{Bevis:2007gh}
 N.~Bevis, M.~Hindmarsh, M.~Kunz and J.~Urrestilla,
 Phys.\ Rev.\ Lett.\ {\bf 100} (2008) 021301.
 [arXiv:astro-ph/0702223].


\bibitem{Hindmarsh:1994re}
 M.~B.~Hindmarsh, T.~W.~B.~Kibble,
 Rept.\ Prog.\ Phys.\ {\bf 58} (1995) 477-562.
 [hep-ph/9411342].

\bibitem{VilShe94}
A.~Vilenkin and E.P.S.~Shellard, 
``Cosmic Strings and Other Defects,'' 
(Cambridge Univ.\ Press, Cambridge, 1994)


\bibitem{Pogosian:1999np}
 L.~Pogosian, T.~Vachaspati,
 Phys.\ Rev.\ {\bf D60 } (1999) 083504.
 [astro-ph/9903361].

\bibitem{Battye:2010xz}
 R.~Battye, A.~Moss,
 Phys.\ Rev.\ {\bf D82} (2010) 023521. 
 [arXiv:1005.0479 [astro-ph.CO]].

\bibitem{Hill:1987qx}
 C.~T.~Hill, H.~M.~Hodges, M.~S.~Turner,
 Phys.\ Rev.\ {\bf D37} (1988) 263.

\bibitem{Jeannerot:2006jj}
 R.~Jeannerot, M.~Postma,
 JCAP {\bf 0607 } (2006) 012.
 [hep-th/0604216]. 
 
 
\bibitem{StringSims}
 C.~Ringeval, M.~Sakellariadou and F.~Bouchet,
 JCAP {\bf 0702} (2007) 023.
 [arXiv:astro-ph/0511646].
\bibitem{Moore:2001px}
 J.~N.~Moore, E.~P.~S.~Shellard and C.~J.~A.~Martins,
 Phys.\ Rev.\ {\bf D65} (2002) 023503.
 [arXiv:hep-ph/0107171].
\bibitem{Hindmarsh:2008dw}
 M.~Hindmarsh, S.~Stuckey and N.~Bevis,
 Phys.\ Rev.\ {\bf D79} (2009) 123504.
 [arXiv:0812.1929 [hep-th]].

\bibitem{Siemens:2006yp}
 X.~Siemens, V.~Mandic and J.~Creighton,
 Phys.\ Rev.\ Lett.\ {\bf 98} (2007) 111101.
 [arXiv:astro-ph/0610920].

\bibitem{Vincent:1997cx}
 G.~Vincent, N.~D.~Antunes and M.~Hindmarsh,
 Phys.\ Rev.\ Lett.\ {\bf 80} (1998) 2277-2280.
 [arXiv:hep-ph/9708427].

\bibitem{Bhattacharjee:1998qc}
 P.~Bhattacharjee and G.~Sigl,
 Phys.\ Rept.\ {\bf 327} (2000) 109-247. 
 [arXiv:astro-ph/9811011].

\bibitem{CRconstraints}
 R.~J.~Protheroe, T.~Stanev,
 Phys.\ Rev.\ Lett.\ {\bf 77 } (1996) 3708-3711.
 [astro-ph/9605036].
\bibitem{Bhattacharjee:1997in}
 P.~Bhattacharjee, Q.~Shafi, F.~W.~Stecker,
 Phys.\ Rev.\ Lett.\ {\bf 80 } (1998) 3698-3701.
 [hep-ph/9710533].
\bibitem{Sigl:1998vz}
 G.~Sigl, S.~Lee, P.~Bhattacharjee and S.~Yoshida,
 Phys.\ Rev.\ D {\bf 59} (1999) 043504.
 [arXiv:hep-ph/9809242].
\bibitem{Wichoski:1998kh}
 U.~F.~Wichoski, J.~H.~MacGibbon, R.~H.~Brandenberger,
 Phys.\ Rev.\ {\bf D65 } (2002) 063005.
 [hep-ph/9805419].


\bibitem{Ellis:2010kf}
 J.~Ellis and K.~A.~Olive,
 arXiv:1001.3651 [astro-ph.CO]



\bibitem{Gherghetta:1999sw}
 T.~Gherghetta, G.~F.~Giudice, J.~D.~Wells,
 Nucl.\ Phys.\ {\bf B559 } (1999) 27-47.
 [hep-ph/9904378].

\bibitem{Moroi:1999zb}
 T.~Moroi, L.~Randall,
 Nucl.\ Phys.\ {\bf B570 } (2000) 455-472.
 [hep-ph/9906527].



\bibitem{Baer:2010kd}
 H.~Baer, R.~Dermisek, S.~Rajagopalan {\it et al.},
 JCAP {\bf 1007 } (2010) 014.
 [arXiv:1004.3297 [hep-ph]].


\bibitem{AbdusSalam:2009tr}
 S.~S.~AbdusSalam, B.~C.~Allanach, M.~J.~Dolan {\it et al.},
 Phys.\ Rev.\ {\bf D80 } (2009) 035017.

\bibitem{Jeannerot:1999yn}
 R.~Jeannerot, X.~Zhang, R.~H.~Brandenberger,
 JHEP {\bf 9912 } (1999) 003.
 [hep-ph/9901357].

\bibitem{DMfromStrings}
 Y.~Cui, D.~E.~Morrissey,
 Phys.\ Rev.\ {\bf D79 } (2009) 083532.
 [arXiv:0805.1060 [hep-ph]].
\bibitem{Bi:2009am}
 X.~-J.~Bi, R.~Brandenberger, P.~Gondolo {\it et al.},
 Phys.\ Rev.\ {\bf D80 } (2009) 103502.
 [arXiv:0905.1253 [hep-ph]].


\bibitem{Sakharov:1967dj}
 A.~D.~Sakharov,
 Pisma Zh.\ Eksp.\ Teor.\ Fiz.\ {\bf 5 } (1967) 32-35.

\bibitem{Fukugita:1986hr}
 M.~Fukugita and T.~Yanagida,
 Phys.\ Lett.\ {\bf B174} (1986) 45. 

\bibitem{Ibe:2004tg}
 M.~Ibe, R.~Kitano, H.~Murayama {\it et al.},
 Phys.\ Rev.\ {\bf D70 } (2004) 075012.
 [hep-ph/0403198].

\bibitem{Giudice:2008gu}
 G.~F.~Giudice, L.~Mether, A.~Riotto {\it et al.},
 Phys.\ Lett.\ {\bf B664 } (2008) 21-24.
 [arXiv:0804.0166 [hep-ph]].

\bibitem{Jeannerot:1996yi}
 R.~Jeannerot,
 Phys.\ Rev.\ Lett.\ {\bf 77 } (1996) 3292.
 [hep-ph/9609442].

\bibitem{Jeannerot:2005ah}
 R.~Jeannerot, M.~Postma,
 JCAP {\bf 0512 } (2005) 006.
 [hep-ph/0507162].
 
\bibitem{Sahu:2005vu}
 N.~Sahu, P.~Bhattacharjee, U.~A.~Yajnik,
 Nucl.\ Phys.\ {\bf B752 } (2006) 280-296.
 [hep-ph/0512350].

 
\bibitem{Affleck:1984fy}
 I.~Affleck and M.~Dine,
 Nucl.\ Phys.\ {\bf B249} (1985) 361.

\bibitem{Davis:1997bs}
 S.~C.~Davis, A.~C.~Davis and M.~Trodden,
 Phys.\ Lett.\ {\bf B405} (1997) 257.
 [arXiv:hep-ph/9702360].

\end{thebibliography}
\end{document}